\documentclass[twocolumn]{aastex631}

\newcommand{\bvf}{Brunt--V\"ais\"al\"a}

\usepackage{tabularx}
\usepackage{amsmath}
\usepackage{xcolor}
\usepackage{float}

\begin{document}

\title{The Impact of a Magnetic Field on stochastically generated g-modes}

\author[0000-0002-2306-1362]{T. M. Rogers} 
\affiliation{School of Mathematics, Statistics and Physics, Newcastle University, Newcastle upon Tyne, NE1 7RU, UK}
\affiliation{Planetary Science Institute, Tucson, AZ, USA}
\email{tamara.rogers@newcastle.ac.uk}

\author{L. M. Corrigan}
\affiliation{School of Mathematics, Statistics and Physics, Newcastle University, Newcastle upon Tyne, NE1 7RU, UK}
\author{L. Barrault}
\affiliation{Institute of Science and Technology Austria (ISTA), Am Campus 1, 3400 Klosterneuburg, Austria}
\author{R. P. Ratnasingam}
\affiliation{School of Mathematics, Statistics and Physics, Newcastle University, Newcastle upon Tyne, NE1 7RU, UK}

\begin{abstract}

Magnetic fields have been proposed as a mechanism for suppressing low-frequency gravity modes in stellar interiors, providing a means of constraining internal magnetic field strengths. However, previous suppression criteria have largely relied on linear theory. Here, we use two-dimensional nonlinear magnetohydrodynamic (MHD) simulations in a core convecting star to investigate the interaction of gravity waves with magnetic fields of varying strengths and geometries. Using both global and local field geometries, we find that amplitude reduction is gradual and frequency dependent, with low-frequency modes preferentially attenuated while higher-frequency modes can be amplified. In order to achieve significant suppression we require a magnetic field strength at least an order of magnitude larger than predicted by commonly used linear criteria. Its likely that this difference arises because magnetic fields can themselves excite waves through Maxwell stresses, offsetting magnetic attenuation. Our results demonstrate that strong magnetic modifications to wave propagation do not necessarily result in equivalent suppression of the global oscillation spectrum. 
The linear criteria typically used to infer field strengths based on mode suppression does not capture all relevant physics and should be used with care. 


\end{abstract}



\section{Introduction} \label{sec:intro}

Magnetic fields are ubiquitous in stellar astrophysics, being either frozen into a star's interior during formation (a 'fossil' field) or actively generated by dynamo action \citep{parker1955, featherstone2009, stello2016}.
In intermediate mass and massive main sequence stars with convective cores, a dynamo occurs within the convective core and the radiative region is then threaded by the fossil field or possibly regenerated locally by e.g. the Tayler-Spruit dynamo \citep{spruit2002,petitdemange2024, Denissenkov2007}.  If the field is fossil in nature it is likely predominantly dipolar, since diffusion would have would act on the smallest scales first, though smaller scale fields could still exist.

During the main-sequence, these fields can significantly influence a star's internal structure and evolution, affecting the rotation profile \citep{spruit2002}, chemical stratification \citep{maeder2005}, convective overshoot \citep{augustson2016} and surface abundances \citep{folsom2012}. Therefore, internal magnetic fields are likely very important in stellar evolution.  Unfortunately, we have little observational constraints on fields in these stars, with large scale stable fields having only been observed in $\approx 10\%$ of massive stars \citep{wade2014,grunhut2014,bowman2022}. Observational constraints on deeper magnetic fields would therefore, be very useful and are just becoming available \citep{vander25,takata26,ihallaine2026}. 

Internal gravity waves (IGWs) are also present in stably stratified stellar radiative regions, such as the envelopes of intermediate mass/massive stars \citep{press1981, kumar1999, rogers2013}. After generation at core-radiative boundaries (CRBs) they are able to affect angular momentum redistribution \citep{zahn1997,talon2004,fuller2014} and chemical mixing  \citep{garcia1991,rogers2017,varghese2023}.  These waves also give rise to standing gravity modes (g-modes) which can be used by asteroseismologists to probe the deep interior and infer quantities such as core mass \citep{aerts2010, aerts2021} near core rotation rate \citep{vanreeth2016} and chemical stratification \citep{miglio2008,pedersen21}. 

Of course, rotation is also a significant factor in stars which can affect both propagating waves \citep{rogers2013, andre2019, varghese2024} and the magnetic field \citep{moss1992, quentin2018}. In rotating stars, the coriolis force is able to lift the degeneracy between modes of different azimuthal order $m$, resulting in frequency splitting which is dependent on $m$ \citep{unno1989, aerts2010}.

In addition to g-modes, more evolved stars such as red giants often present mixed modes, which display characteristics of both gravity and pressure modes in different regions of the star. These mixed modes then provide a powerful diagnostic for the structure of both the core and envelope \citep{beck2011,bedding2011,deheuvels2012}, and could be used to investigate both the strengths and geometries of internal magnetic fields \citep{mathisbugnet21,deheuvels2023,li2023,hatt2024,villate2026}. 

With both magnetic fields and IGW co-existing in stellar radiative interiors, it is likely that their interactions hold important implications for the interior dynamics and evolution of the star.  These interactions have been explored in a range of contexts, including in solar type stars \citep{schatzman1993, barnes1998, vigeesh2017, amard2026} and red giants \citep{loi2018,rui2023,david2026}. In the context of solar type stars,  \cite{rogers2009,rogers2010} showed that the presence of a toroidal field present at the solar tachocline could result in the reflection, trapping and filtering of IGW which could have significant implications for transport, mixing and energy redistribution in the radiative interior.

More recently in red giants, it has been suggested that sufficiently strong core magnetic fields can significantly suppress dipole ($l=1$) g-modes by trapping and dissipating gravity waves in the core as part of a 'magnetic greenhouse effect' \citep{fuller2015,rui2023}. 
This is supported by the work of \cite{stello2016}, which found strong suppression of dipolar modes and notable suppression of quadrupolar modes in Kepler observations of red giants. Building on this, \cite{lecoanet2016} used linear theory and numerical simulations of a single propagating wave to propose that IGW encountering strong magnetic fields undergo mode conversion to magnetic Alfv\'en-like waves, preventing their reflection and inhibiting the formation of g-modes. 

However, these interpretations are still under debate. 
\cite{mosser2017} showed that even strongly depressed gravity-dominated dipole modes retain mixed-mode characteristics and evenly-spaced periods, suggesting that oscillation modes cannot be fully suppressed in the radiative core of red giants. In addition, \cite{loi2020} suggested that magneto-gravity waves are not entirely trapped within the core, and that some are able to reflect out of the core much like pure gravity waves. From an asteroseismic point of view the impact of a magnetic field on g-modes is still rather uncertain due to its complexity and various assumptions made in the previous treatments, such as simplified field geometry or highly localized field interaction.  

Magnetic fields can affect IGWs both at generation and during propagation. \cite{eltayeb1970,cattaneo2003} showed that magnetic fields are able to affect convective velocities which could in turn impact generated wave amplitudes.  During propagation IGW can bend and distort magnetic field lines, which then exert a Lorentz force back onto the waves. This additional restoring force could alter wave velocities \citep{rogers2010,fuller2015,lecoanet2016} and therefore affect both the efficiency of angular momentum redistribution and mixing of chemical species (Corrigan et al., under review). In turn, this could also alter the dynamics and strength of wave-driven flows and, therefore, the differential rotation profile within the star.

Further, the presence of a magnetic field introduces an additional Alfv\'en frequency term into the waves' dispersion relation \citep{schatzman1993,macgregor2011}, indicating the field is capable of affecting IGW frequencies in a way which depends on the field strength. 

These effects extend to standing g-modes, with \citep{goughthompson90,goode1992,hasan1992,hasan2005,mathisbugnet21}  proposing that magnetic fields can cause modes to be shifted in frequency. In rotating stars, \citep{goode1992,dziembowski1996,mathisbugnet21,bugnetmathis23} also suggested that fields are able to produce asymmetries in rotationally split multiplets by introducing different magnetic shifts between components of different $|m|$. 

In this work we present results from 2D magnetohydrodynamic (MHD) simulations of interactions between magnetic fields and convectively generated g-modes. By exploring a range of field strengths and geometries we assess the impact of the magnetic field on g-mode amplitudes and frequencies.  These simulations build on previous work by self-consistently simulating the generation of a full spectrum of waves, which then interact with a time-dependent magnetic field with all back-reactions and nonlinear effects accounted for.  

\section{Numerical Method} \label{sec:numerics}

The numerical method used here is described in \cite{rogers2013} for the hydrodynamic model and \cite{rogers2010} for the magnetic model.  In summary, we solve the MHD equations in two dimensions (2D) in the anelastic approximation in cylindrical geometry, representing an equatorial slice of the star.  The equations are discretized using a spectral fourier expansion of the variables in the horizontal dimension ($\phi$) and a second-order finite difference scheme in the radial (r) direction.  
\begin{figure}[H]
        \centering
        \includegraphics[trim={0.0cm 0.0cm 0 0.0cm},clip,width=\columnwidth]{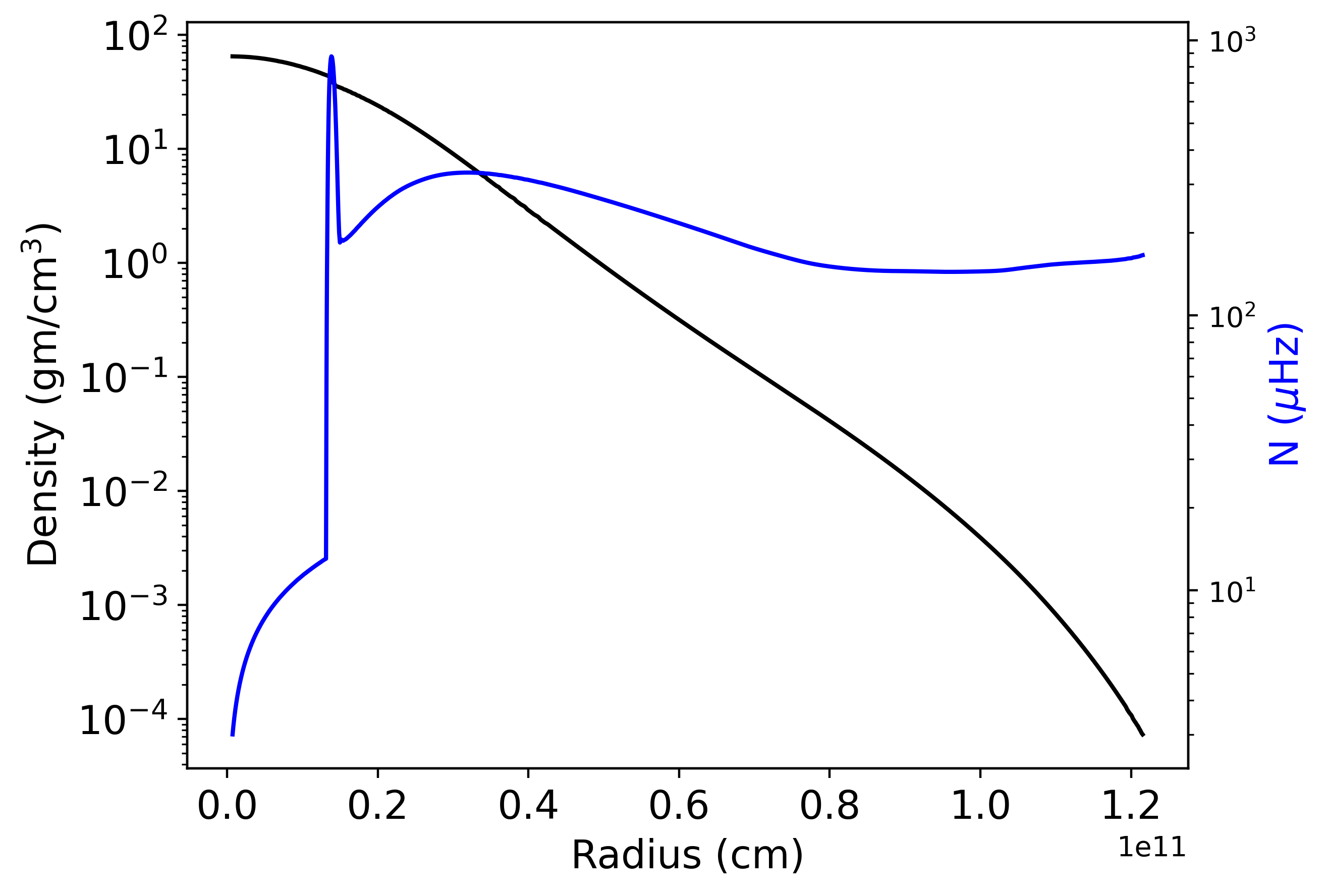}
        \caption{Reference state density, $\overline{\rho}$ (black) and \bvf{} frequency (blue) as a function of radius for the presented simulations. \label{fig:stellar_param}}
        \centering
\end{figure}
The model presented is that of a 2M$_{\odot}$ star at $X_c$=0.5, representing a slightly evolved star on the main sequence. This initial reference state was taken from MESA (version r24.08.01). The 1D model used mixing length theory to treat convective mixing \citep{boehmvitense1958}, an exponential overshooting profile with parameters $f = 0.016, f_0 = 0.008$ \citep{herwig2000} and base metallicity $Z = 0.014$ \citep{asplund2009}. Neither rotation or mass loss were included in the 1D model. From this 1D model we extract reference state quantities such as temperature, density and stratification (see Fig.~\ref{fig:stellar_param}). 

Our simulations handle convective/radiative regions through a super/sub-adiabatic temperature gradient which we compute from the \bvf{} frequency from MESA, for the radiative region.  In the convective region we set the superadiabaticity to a constant, small value ($10^{-6}$) to force convection.  In the radiative zone, the \bvf{} frequency from MESA includes the chemical composition gradient which we then account for with an altered sub-adiabatic temperature gradient.   We resolved the simulation domain from 0.05R* ($\sim 7\times 10^{8}$cm) to 0.93 R* ($\sim 1.2\times 10^{11}$cm).    The reference state \bvf{} frequency and density of the simulation are shown in Fig.~\ref{fig:stellar_param}, where the narrow spike due to the composition gradient is clearly visible. 

All simulations have resolution of 1024 horizontal grid points (corresponding to 722 fourier modes) and 1500 vertical grid points, with 500 zones used for the convection zone including the \bvf{} spike and 1000 used for the radiative region.  The hydrodynamic simulations have fixed viscous and thermal diffusivities which are constant with radius ($=5\times 10^{11}$ cm$^2$ s$^{-1}$), resulting in a Prandtl number of 1.  This then leads to a Reynolds numbers within the convection zone, varying from $\sim$1000-4000, depending on the model, as seen in Table ~\ref{tab:models}.

Our vertical resolution and thermal diffusivity limit the lowest frequency we can adequately resolve ($\kappa k_r^2\sim \omega$) to be that of a wave with radial wavelength of $\sim 10^{7}-10^{8}$cm. For fourier mode $m=1$, this corresponds roughly to radial order $n$ = 15 or approximately 12$\mu$Hz.  This is {\it not} a frequency resolution, but below this frequency, thermal diffusion occurs on a timescale similar to the wave frequency.  Of course, this is an estimate, we see in  Fig.\ref{fig:radstruct} that we can resolve waves with slightly lower frequencies. We note that the value of our thermal diffusivity is higher than it is in the actual star near the core, but lower than it is in the actual star, near the surface.  Given the rather large thermal dissipation in the outer radiative envelope of the actual star, it's likely that thermal diffusion similarly limits resolution of modes much below this frequency in actual stars as well, which results in broadening of modes, particularly modes which are not coherently driven.   Moreover, even at higher frequencies thermal diffusion affects the mode width and lifetime. 

All simulations are run for at least 3$\times$10$^{7}$s, corresponding to a minimum frequency resolution of 0.033$\mu$Hz. Some simulations have run much longer. This results in approximately 50-300 convective turnover times (depending on the model), once the field is imposed, where we have defined convective turnover time using the rms velocity and depth of the convection zone, see Fig.~\ref{fig:vrms}. 

We initially run the hydrodynamic simulation for approximately 20 convective turnover times.  We then impose a radial magnetic field by imposing the vector potential $\bf{A}$, where ${\bf B}=\nabla \times {\bf A}$, on the $m=1$ component for the duration of the simulation.  This ensures that the imposed field satisfies the divergence-free condition in this geometry. This imposed field is held fixed in time.
\begin{figure*}[!t]
        \centering
        \includegraphics[trim={0.0cm 0.0cm 0 0.0cm},clip,width=\textwidth]{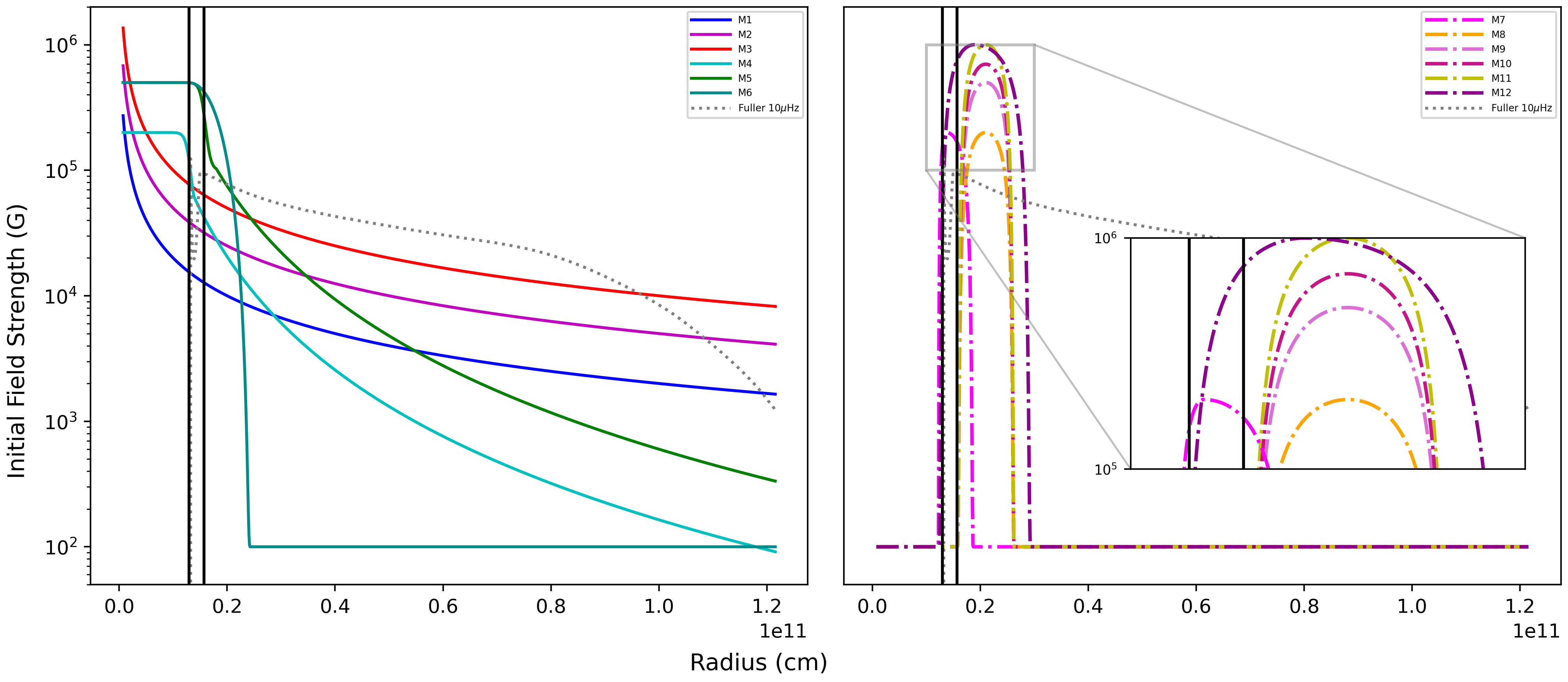}
        \caption{Imposed radial field strength for various magnetic models. "Global" fields are shown with solid lines on the left, while "local" fields are shown with dash-dotted lines on the right. The inset shows a zoomed region around the \bvf{} spike (denoted by the vertical black lines) for the "local" fields. Note that model M5 and M6 overlap within the convection zone, but M6 drops to the "floor" field strength of 100G just outside the convection zone.  The dotted grey line represents the "critical" field strength for a 10$\mu$Hz wave, using Equation (2).}
        \label{fig:initfields}
        \centering
\end{figure*}

We then test a variety of different radial field strengths and geometries, as shown in Fig.~\ref{fig:initfields}.  Models M1-M3 are the most basic configuration with a vector potential that is constant with radius, resulting in a radial field that decays as 1/r. This dependence effectively means that the m=1 toroidal field is then imposed to be zero.  These models were run to most closely mimic the imposed field geometries of \cite{fuller2015} and \cite{lecoanet2016}, which have virtually no toroidal field. 

Models M4-M6 are informed by our 3D dynamo simulations \citep{rathish26}, which have large field strengths in the convection zone, due to the dynamo in that region.  Within the radiative region the field strength drops off dramatically, as we imagine it might in a main sequence star if there is no active dynamo within the radiative region.  Since we have virtually no constraints on the field within the bulk of the radiative zone, we run multiple models varying the strength and radial dependence of the field in this region.   We collectively refer to these models (M1-M6) as "global" magnetic fields (left hand panel of Fig.~\ref{fig:initfields}), by which we mean that there is a strong field in the convection zone and it connects smoothly across the convective-radiative boundary (CRB).  We expect these models to best reflect the magnetic field in a main sequence star such as the one simulated here.

Our second set of models, M7-M12, we refer to as "local" models, meaning that they are highly localized with radius.    One can not impose a completely field-free region, as this would lead to current sheets which could not be resolved.  Moreover, we don't believe there are truly field-free regions in a star.  Still, these highly localized models, shown in the right hand panel of Fig.~\ref{fig:initfields}, have peak field strengths $10^{3}-10^{4}$ times larger than the surrounding field\footnote{Larger peak values and more narrowly localized fields and localized fields in lower \bvf{} environments seem to lead to magnetic instabilities}.   We set the "floor" in these models to be a magnetic field strength of 100G. Although we don't envision these models as particularly realistic for this reference state model, they do allow us to investigate the role of field geometry in a more complete way.  Moreover, because they are not strongly connected to the convection zone, they allow us to isolate the effects of the magnetic field on the generation of waves from its effect on the propagation of waves. They also allow us to explore higher field strengths and lower magnetic diffusivities.  For reference, we also show on Fig.~\ref{fig:initfields} the critical field strength for a 10$\mu$Hz wave, as calculated from the criteria in \cite{fuller2015} (Equation (2) below) but adjusted for this two-dimensional geometry.

For all models, the imposed radial field on $m=1$ results in an effectively imposed toroidal field at $m=1$, due to the divergence-free condition. For models M1-M3, this toroidal field is zero, but for the other models it is non-zero and depends on the radial gradient of the radial field.  Therefore, the highly localized models M7-M12, lead to very high toroidal fields in that region to satisfy this condition. Both the radial and toroidal field that are {\it not} $m=1$ can evolve self-consistently in response to the flow. The radial field is then constant across the domain boundaries, and there is no current at the boundaries.  This imposed geometry ensures $\nabla\cdot{\bf B}=0$ initially and the vector potential formulation ensures it remains zero.  
 
The magnetic diffusivity in most simulations is $5\times 10^{12}$ cm$^2$ s$^{-1}$, resulting in magnetic Prandtl numbers of 0.1.  Models M8-le, M8-lle, M11-le and M11-lle have reduced magnetic diffusivity, resulting in a magnetic Prandtl number of 1 and 5, respectively.  Those models were run to test the dependence of the results on the Lundquist number and will be discussed Appendix~\ref{app:nond}.  The field strengths, as well as some other non-dimensional numbers for the simulations are listed in Table ~\ref{tab:models} and expanded upon in Appendix~\ref{app:nond}.

\begin{deluxetable}{lccc}
\tabletypesize{\small}
\tablewidth{0pt}
\tablecaption{The models and their parameters. Initial magnetic field strength, given as the peak imposed radial field strength within the radiative zone (see Fig.~\ref{fig:initfields}). Reynolds numbers ($Re=UL/\nu$) within the convection zone, calculated using the rms velocities, the depth of the convection zone (1.3$\times 10^{10}$cm) and the viscous diffusion (5$\times10^{11}$). All models have an initial uniform rotation rate of $10^{-6} $ rad/s, so the Rossby number varies between 15-75, depending on the rms velocity. Lundquist numbers are for a 10$\mu$Hz wave at 3$\times 10^{10}$cm, see Appendix ~\ref{app:nond} for radial structure.  The Froude Number, defined as Fr=$\omega/N$ at the \bvf{} spike is approximately 10$^{-2}$ for a 10$\mu$Hz wave.  Radial dependence of Fr for various frequencies is shown in Appendix~\ref{app:nond}.}  \label{tab:models}
\tablehead{
\colhead{Model}  & \colhead{$B_r{}_{\rm pRZ}$} (G) & \colhead{Re} &\colhead{Lu (10)}}
\startdata
F              & 0 & 3900 &NA\\
M1            & $1 \times 10^{4}$ & 2600 & 200 \\
M2            & $4 \times 10^{4}$ & 1300 & 200 \\
M3           & $8 \times 10^{4}$ & 1300 & 200 \\
M4            & $2 \times 10^{5}$ & 1300 & 200\\
M5            & $5 \times 10^{5}$ & 780 & 200\\
M6             & $5 \times 10^{5}$& 780 & 200\\
M7            & $2 \times 10^{5}$& 3900 & 200\\
M8           & $2 \times 10^{5}$& 3900 & 200\\
M8-le           &$1\times 10^{6}$ & 3900 & 2000\\
M8-lle           &$1\times 10^{6}$ & 3900 & $10^{4}$ \\
M9           & $5 \times 10^{5}$& 3900 & 200\\
M10          & $7\times 10^{5}$& 3900 & 200 \\
M11             & $10^{6}$&3900 & 200 \\
M11-le           & $10^{6}$&3900 & 2000 \\
M11-lle          & $10^{6}$&3900 & $10^{4}$ \\
M12               & $10^{6}$ &3900 & 200\\
\enddata
\end{deluxetable}

\label{tab:models}
Although this field geometry (typically referred to as an equatorial dipole) is  artificial, it mimics well  the dynamical interaction between IGW, which are mostly horizontal, with a (largely) radial (poloidal) magnetic field.   Moreover, although the radial and toroidal field are artificially imposed and fixed in time for $m$ = 1, the radial and toroidal field that are not $m$=1 are allowed to evolve freely.   Therefore, this set up allows time dependence in which the flow alters the field and the field reacts back on the flow, this results in a toroidal and radial field which are both time dependent.  
\begin{figure}[H]
        \centering
        \includegraphics[trim={0.0cm 0.0cm 0 0.0cm},clip,width=\columnwidth]{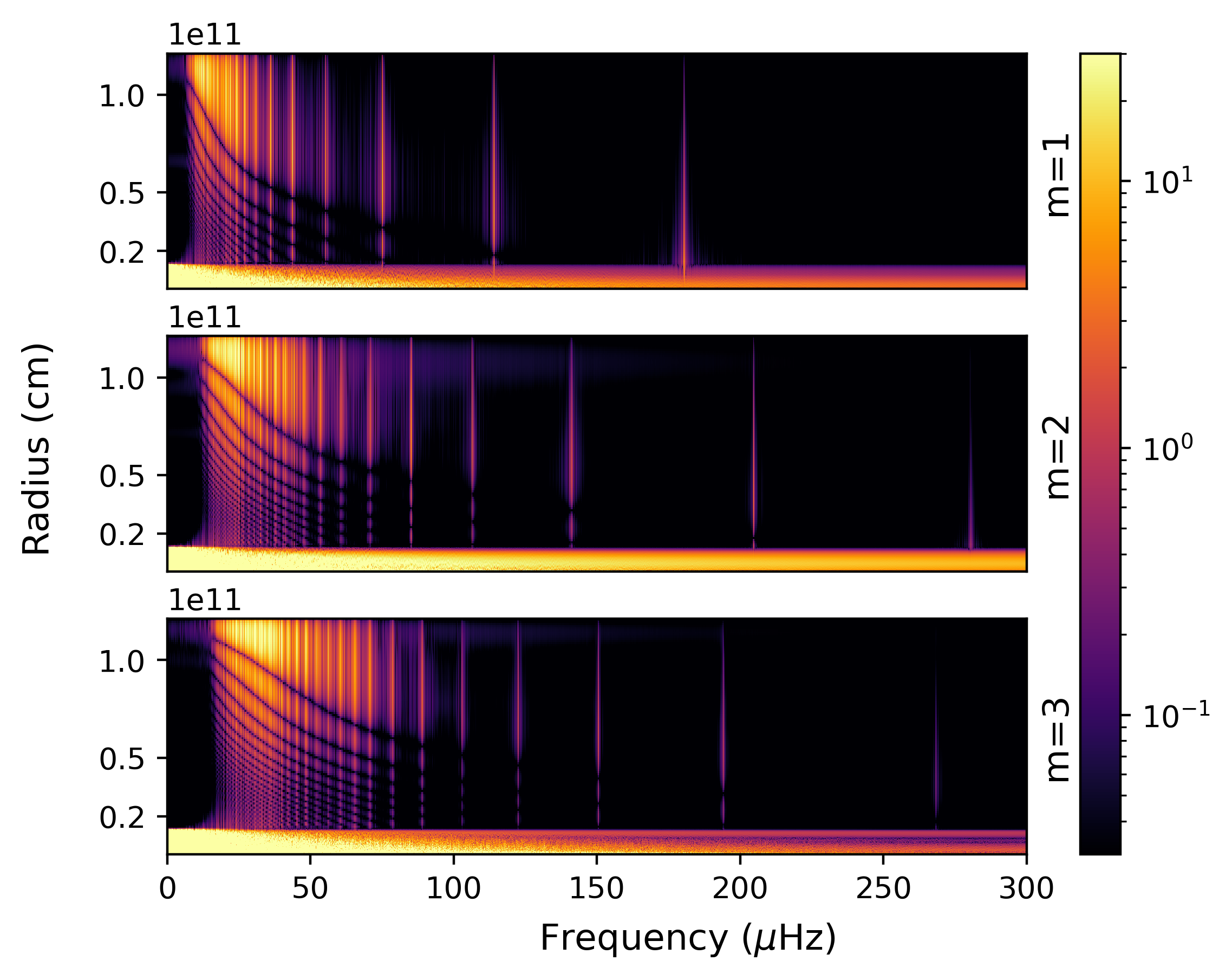}
        \caption{Radial velocity as a function of frequency and radius for the fiducial hydrodynamic model (F), for horizontal wavenumbers m=1,2,3. \label{fig:hydrospec}}
        \centering
\end{figure}


\section{Results} \label{sec:results}

\subsection{Initial Hydrodynamic Model}
As described in Sec.~\ref{sec:numerics} the initial hydrodynamic model is a 2M$_{\odot}$ at X$_c$=0.5 with $\Omega=10^{-6}$ rad/s (our "fiducial" Model F in Table 1). The frequency spectra as a function of radius for Model F is shown in Fig.~\ref{fig:hydrospec}.  There we show the colormap of radial velocity as a function of frequency and radius for the three lowest horizontal wavenumbers.  Looking at this figure we see hallmark features of standing g-modes. First, radial order increases with decreasing frequency, as expected from the basic dispersion relation.  Second, larger horizontal wavenumber (smaller horizontal scales) have higher frequencies at a given radial wavenumber.  Finally, larger horizontal wavenumbers are dissipated more strongly at the shortest frequencies \citep{kumar1999}. In addition to resolving many of the asymptotic (low frequency) g-modes, the simulations also recover the low order modes, including the f-mode, which do not satisfy the assumptions of JWKB theory \citep{BenderOrszag1999,unno1989}.  

\begin{figure}[H]
        \centering
        \includegraphics[trim={0.0cm 0.0cm 0 0.0cm},clip,width=\columnwidth]{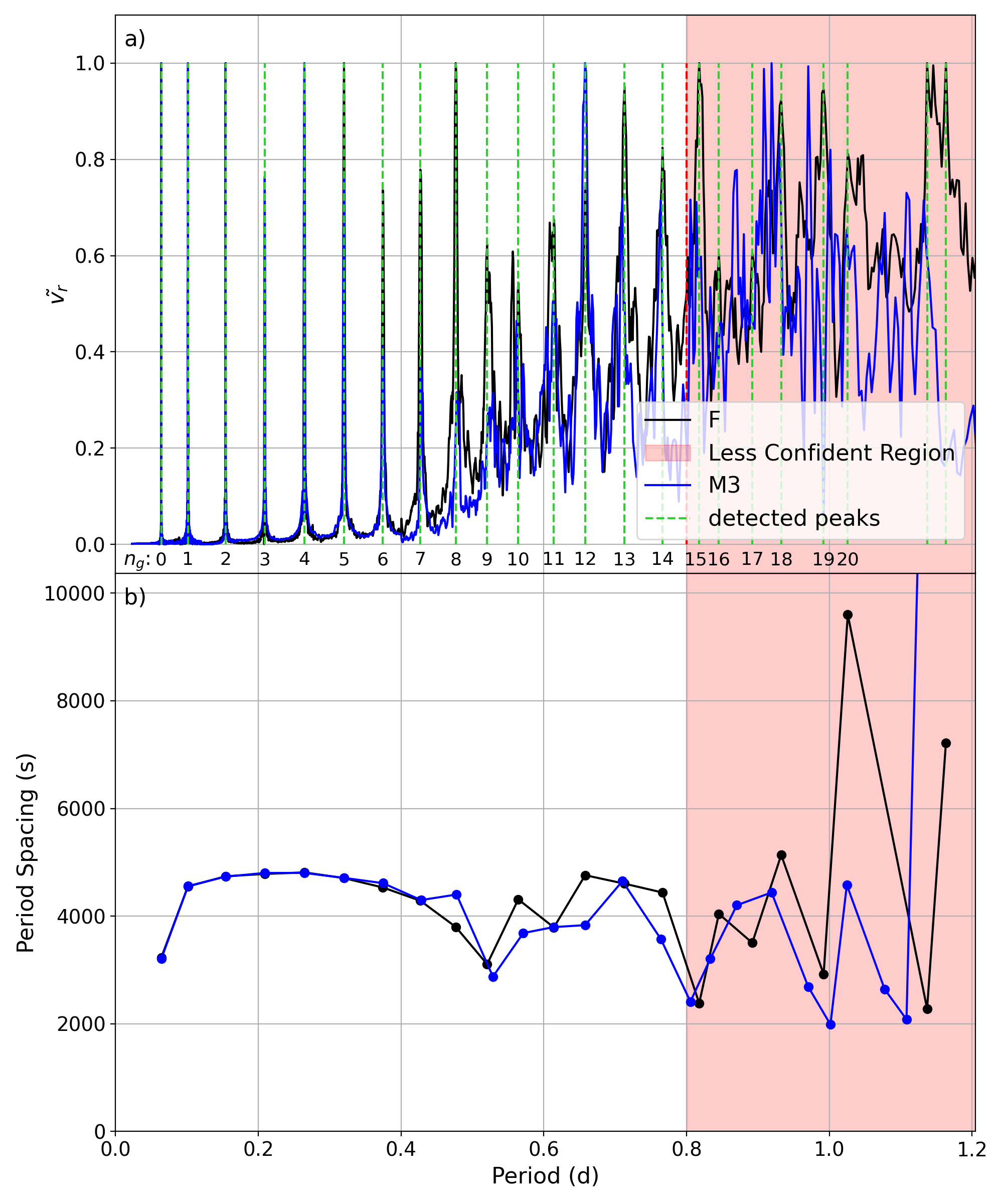}
        \caption{(a) Radial velocity as a function of period for the fiducial, slow rotating, hydrodynamic model F (black) alongside the magnetic model M3 (blue) for m=1 with radial order indicated. (b) Period spacing diagram for the same model, showing a relatively flat period spacing at the shortest periods and variation at longer periods where identifying modes becomes more difficult.   \label{fig:periodspacing-fid}}
        \centering
\end{figure}

We show the wave amplitudes as a function of period alongside the period spacing diagram (difference in period for consecutive radial orders) for Models F and M3 in Fig.~\ref{fig:periodspacing-fid}.  This figure is a composite radial spectrum obtained in the following way. We obtain radial velocity for a given horizontal wavenumber at every radius.  Low frequency modes are more visible near the surface due to the density stratification increasing their amplitude. Conversely, high frequency modes are clearer at low radii.  It is therefore easier to use a "composite" spectra, where we combine sections of the frequency spectrum from different radii according to where they are most visible.  These selected intervals are then normalized by dividing by their maximum value and combined to yield this "composite spectrum".  This is a valid way to identify standing mode frequencies since the frequencies are independent of radius, as seen in Fig.~\ref{fig:hydrospec}.   To further improve peak-identification we use a Savitzky-Golay filter with a small window size to suppress noise everywhere without smoothing over established peaks.  This smoothing is not applied to the lowest frequencies where peaks are much closer together (and thermal diffusion is smearing the line).  Standing modes are then identified as peaks in the composite spectrum using SciPy's find peaks routine \citep{scipy2020}.  

\begin{figure}
        \centering
        \includegraphics[trim={0.0cm 0.0cm 0 0.0cm},clip,width=\columnwidth]{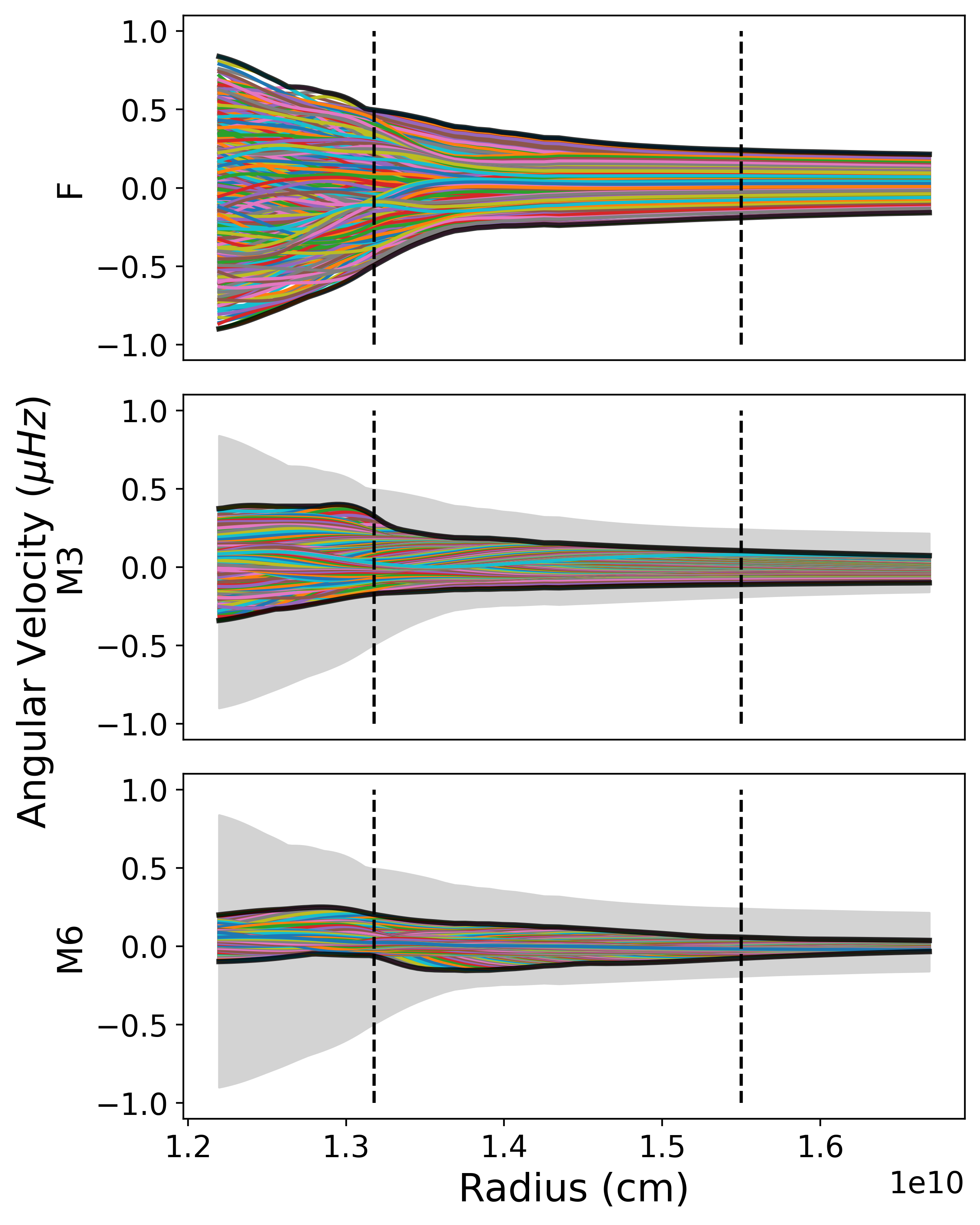}
        \caption{Time variability of the angular velocity for non-magnetic Model F alongside the magnetic models M3 and M6. Each line represents the horizontally averaged angular velocity, with time intervals of 1000s, between 2.5$\times 10^{7}$ and 3.5$\times 10^{7}s$. The magnetic models are shown with a grey shaded envelope, representing the angular velocity of the fiducial model for comparison.  We see that the time variability is drastically reduced in the magnetic cases, alongside the overall radial differential rotation across the BV spike (indicated by vertical black lines). While we only show models F, M3 and M6, the trend is generally the same for all magnetic models. In general the time variability is decreased by a factor of two or more and the differential rotation is reduced by $\sim 20-50$\%.  The time variability is of order $\sim 1\mu$Hz in hydrodynamic models and $\sim 0.3 \mu$Hz in magnetic models. The differential rotation is of order $\sim 0.2\mu$Hz in hydrodynamic models and $\sim 0.1\mu$Hz in magnetic models.  These are rough ranges due to the large  variability within and across models. \label{fig:drvariability-rot}
        }
        \centering
\end{figure}

In Fig.~\ref{fig:periodspacing-fid} we see a fairly flat period spacing at the largest radial wavelengths/lowest frequencies ,where the asymptotic result for high radial order dipole g-modes of $\Delta P =\Pi_{0}/\sqrt{2}$ doesn't necessarily apply. At higher periods, better described by the asymptotic regime, the period spacing becomes irregular with dips and a possible change in slope.   These dips and slope may be reflective of the spike in the \bvf{} frequency and/or differential rotation.  Unfortunately, at higher periods, the peaks become harder to identify due to thermal diffusion and noise around the modes (see Fig.~\ref{fig:drvariability-rot}).  Therefore, interpreting the dips and shape as reflective of the physical conditions may not be reliable.  We see no consistent difference in the period spacing for the magnetic model.

In addition to thermal diffusion broadening the modes, differential rotation across the CRB and time variability of that differential rotation also contribute to "noise" around modes. Figure ~\ref{fig:drvariability-rot} shows the differential rotation and time variability of differential rotation across the CRB for Model F, Model M3, and Model M6. There we see that for the fiducial model the differential rotation across the CRB is approximately 0.5$\mu$Hz and the time variability is 2$\mu$Hz over the time in which spectra are collected.  We see that this variability is significantly reduced in the magnetic cases, proportional the the radial magnetic field strength across the CRB. This is true for all our models when the field is continuous across the CRB, as expected. 

\begin{figure*}[!t]
        \centering
        \includegraphics[trim={0.0cm 0.0cm 0 0.0cm},clip,width=0.8\textwidth]{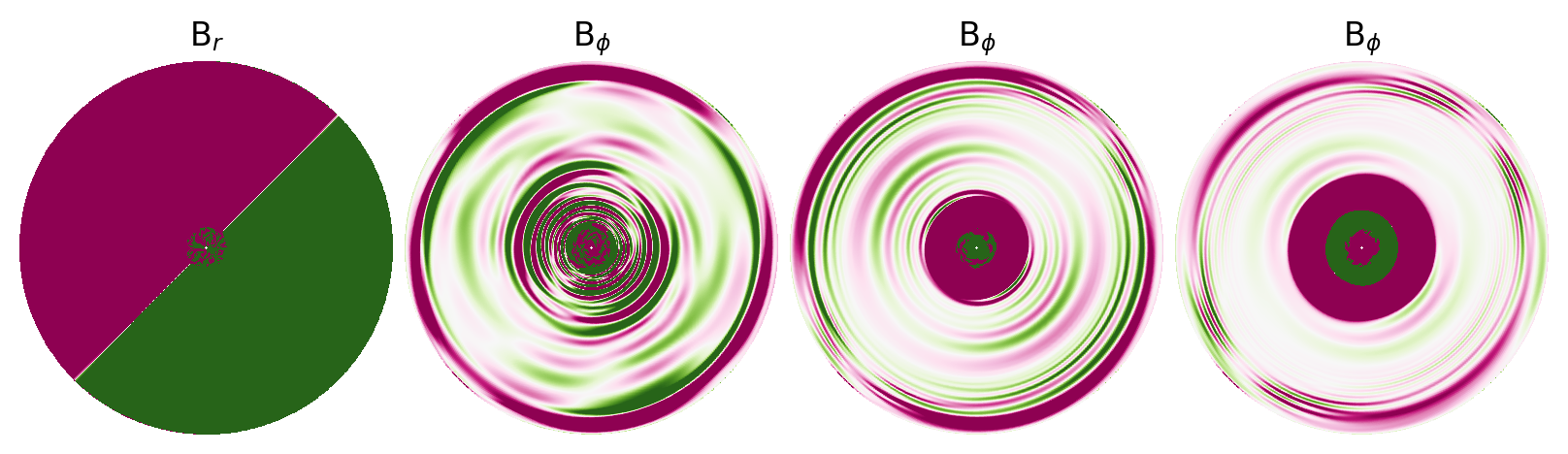}
        \caption{Time snapshots of the radial field (early in the simulation, left hand column) and the toroidal field for Model M3 as time progresses rightward.  Early on toroidal field is induced by IGW, but as time goes on the wave amplitudes are decreased and the induced toroidal field is also reduced.  Green/Magenta are saturated at $\pm$100G. \label{fig:bfieldevolv}}
        \centering
\end{figure*}

\subsection{Evolution of the Imposed Field}\label{sec:fieldevol}

A small differential rotation between the convective core and radiative region is self-consistently induced in all models.  This causes the initially radial field to be sheared into strong toroidal field at the CRB (at $m\ne1$). The induced toroidal field grows in time and spreads into the radiative region, as seen in Fig.~\ref{fig:bfieldevolv}, which shows time snapshots of the radial and toroidal field throughout the domain.   In the strong field cases the mean differential rotation is reduced, but still not eliminated, and at that point the toroidal field diffuses further into the radiative region. Further into the  radiative zone only very weak toroidal field is induced by the IGW themselves.  As time goes on the IGW amplitudes reduce due to the presence of the magnetic field (see Sec.~\ref{sec:fieldamp}) and therefore the induced toroidal field throughout the radiative region is also reduced (see Fig.~\ref{fig:bfieldevolv} and Fig.~\ref{fig:bfieldosc}).

\begin{figure}[!hbtp]
        \centering
        \includegraphics[trim={0.0cm 0.0cm 0 0.0cm},clip,width=\columnwidth]{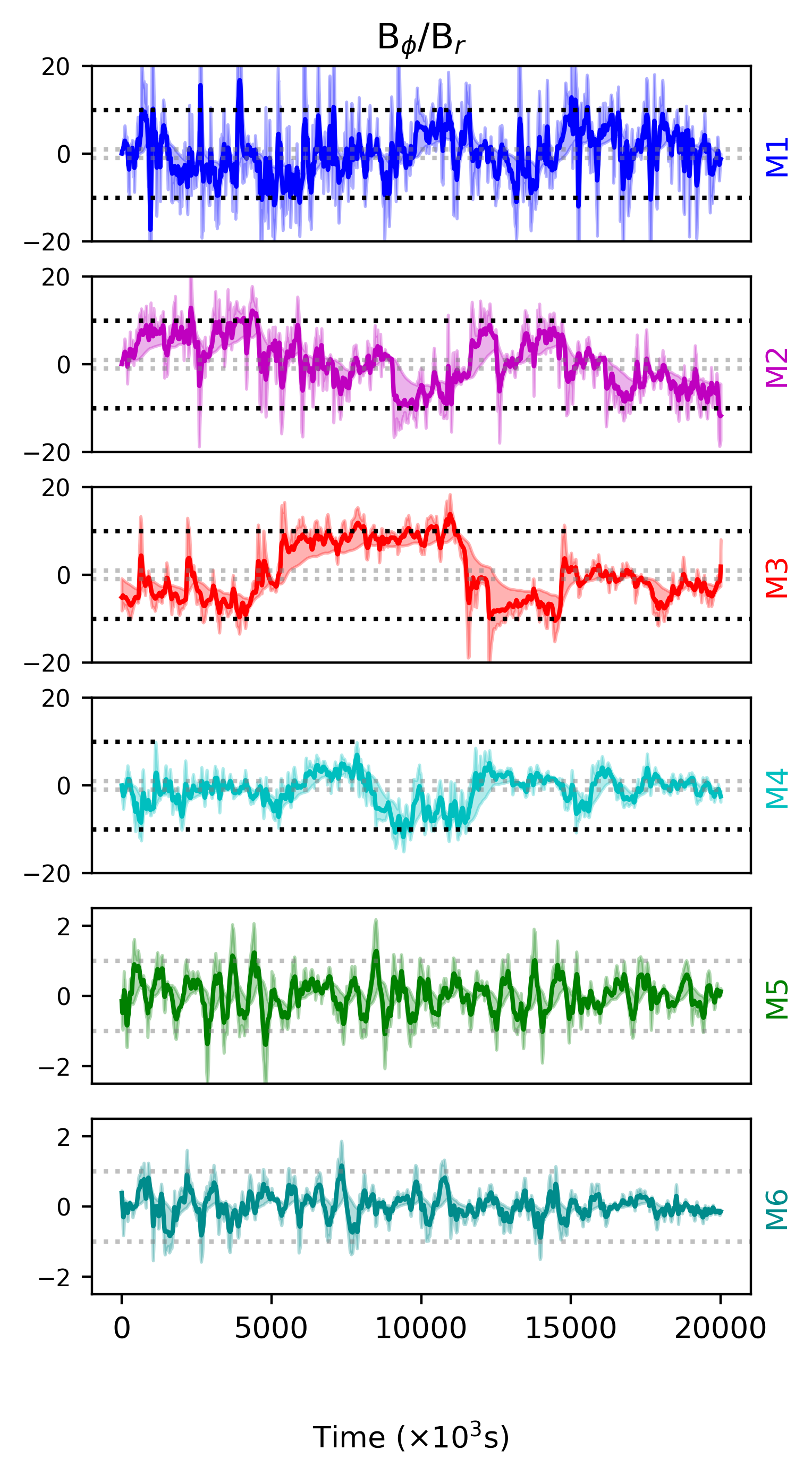}
        \caption{Ratio of the toroidal to radial field in the \bvf{} spike.  The colored lines show the  mean over the spike while the shading shows the range of variation.  Horizontal black dotted lines indicate a ratio of $\pm$10, while horizontal grey dotted lines indicate a ratio of $\pm$1.  Note the different y scale for model M5 and M6.}\label{fig:brat}
        \centering
\end{figure}

An important factor for (magneto-)asteroseismology is the ratio of the toroidal to poloidal field strength in the region of \bvf{} spike.  We show this ratio for each of the global models (M1-M6) in Fig.~\ref{fig:brat}. There we see a few interesting features. First the ratio is generally significantly larger than one (denoted by grey horizontal lines) except in the strongest field cases.  Second, the sign changes regularly. Since the radial field is virtually fixed, this represents flipping of the sign of the (mean) toroidal field.  Third, the ratio is larger for weaker imposed radial field, reaching values of $~\sim$20 for the weakest imposed field, but hovering $\sim\pm$1-2 for the largest field strength.  Moreover, the weakest field also shows the largest fluctuations in this ratio.  If this trend continued, we would expect even larger ratios in weaker field cases.  We note that even our weakest field model, M1, still has a higher radial field strength at the CRB than that inferred in \cite{takata26} or found in \cite{rathish26}.  Therefore, this trend is consistent with those measured values. 

Although the weakest field cases have the largest ratio of toroidal to poloidal field, they still have the weakest absolute toroidal field strength. Generally, the toroidal field at the CRB ranges from $\sim10^{5}$G up to 1MG.  In fact across all global models the toroidal field strength only varies by a factor of $\sim 2$, despite the radial field varying by more than an order of magnitude in that region.  Of course, this is likely to be limited by our low magnetic Reynolds number (high magnetic diffusion) in the region, so it is possible that the toroidal field strengths in stars are even higher. As one can see in Fig.\ref{fig:drvariability-rot} the differential rotation across the CRB is significantly reduced in Model M6 (the same is true for M5).  This results in a reduction of induced toroidal field as time progresses, as seen in Fig.~\ref{fig:brat}.  Strong radial field at the CRB reduces differential rotation, thereby reducing the strength of the induced toroidal field, as expected.

\begin{figure}[ht]
        \centering
        \includegraphics[trim={0.0cm 0.0cm 0 0.0cm},clip,width=\columnwidth]{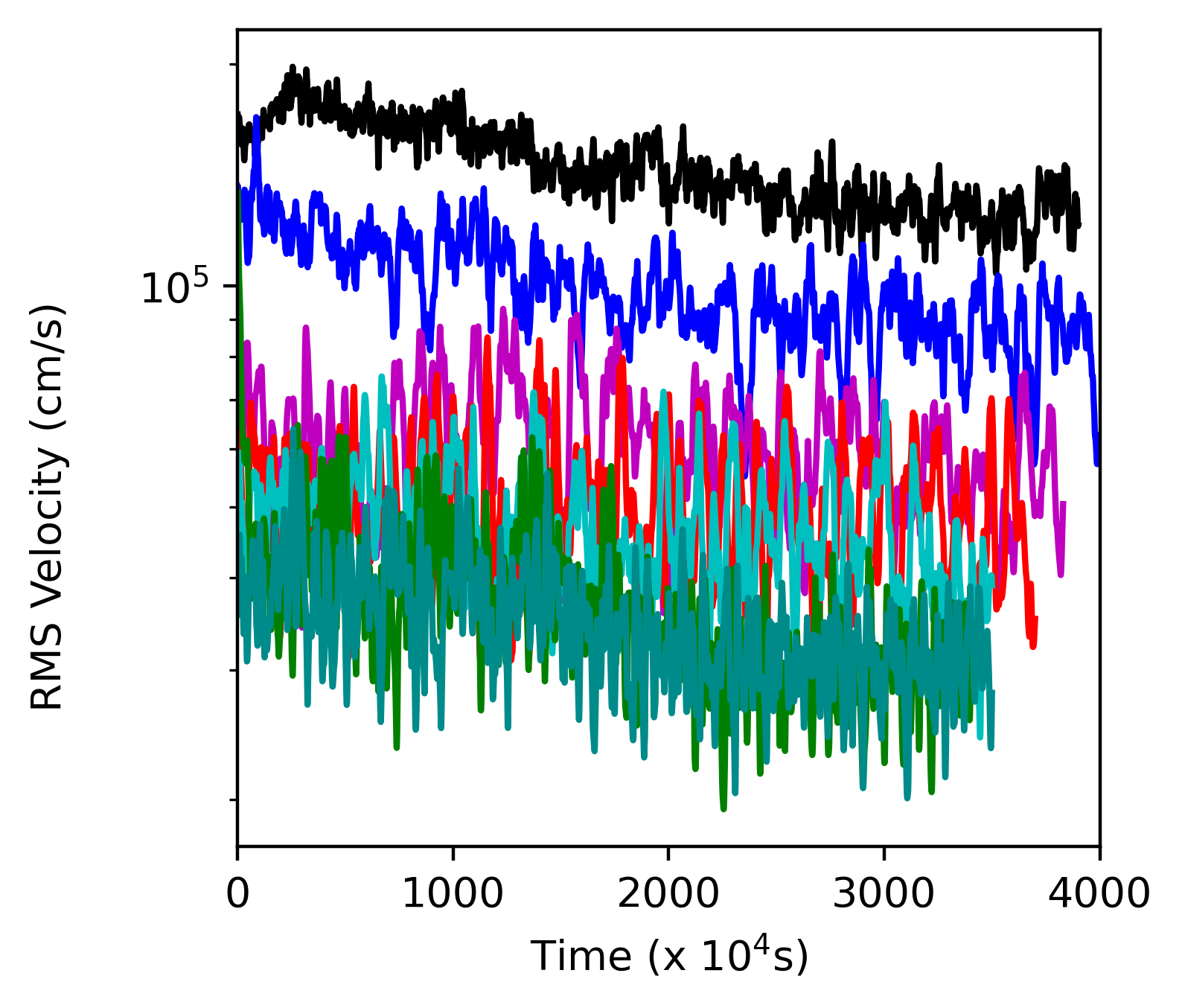}
        \caption{(a) Root mean squared (RMS) velocity as a function of time for the hydrodynamic model (black) and its magnetic counterparts (colors as in Fig.~\ref{fig:brat}). The higher core magnetic field strengths show significantly lower rms velocities (a factor of three for the largest field strengths) than their hydrodynamic counterpart. } \label{fig:vrms}
        \centering
\end{figure}

In all cases the introduction of a magnetic field reduces the root mean square velocities within the convection zone, as seen in Fig.~\ref{fig:vrms}.  This is clearly dependent on the field strength, with Model M1 reduced far less than Model M5. Although there is a clear trend of decreasing rms velocity with increasing field strength, in models M2-M4 the time variation of the rms velocity is larger than the variation due to the magnetic field strength. Moreover, the presence of the magnetic field increases the variability of the rms velocity compared to the hydrodynamic counterpart, which has very little fluctuation. The reduction in rms velocities alone will have an impact on the amplitudes of the waves generated at the CRB.  The "localized" fields M7-M12 have no impact on the convective velocities.

\subsection{Impact of the Magnetic Field on the g-mode Amplitudes} \label{sec:fieldamp}

We now turn to investigating how the magnetic field affects the g-mode amplitudes.  Before discussing our results it is worth reviewing the criteria for mode suppression proposed by \cite{fuller2015} and \cite{lecoanet2016} and subsequently used in numerous works to infer internal magnetic field strengths \citep{lecoanet22,rui25wd,ruima26,durfeldt26}.  The original criteria from \cite{fuller2015} was that any frequency lower than the magneto-buoyancy frequency, defined as:

\begin{equation}
\omega_{MB}=\left(\frac{2 B_r^2 N^2}{\pi\rho r^2}\right)^{1/4}
\end{equation}
could not propagate or, conversely, that a field above the critical field strength, defined as: 

\begin{equation}
B_c=\sqrt{\frac{\pi\rho}{2}}\frac{\omega^2 r}{N}
\end{equation}
would suppress modes at that frequency\footnote{In our 2D geometry we must replace $\sqrt \ell(\ell+1)/r$ with $k_h$ \citep{lecoanet2016} resulting in the factor of 2 in (1) and (2) being replaced by 1.}.  In this original picture this suppression was {\it complete}, the g-mode was converted to Alfv\'en waves and completely dissipated.  In this {\it theoretical} definition of mode suppression, the structure of the g-mode is lost and the mode is effectively eliminated.  In the following, we will use this "strong" suppression criteria to mean {\it both} that the amplitude is reduced by at least an order of magnitude and that the mode structure is disrupted. This will be distinguished from a "weak" criteria where there is just amplitude reduction of the mode, while retaining the g-mode structure. We will show that there could be a significant difference in field strength between the two criteria.

\begin{figure}[!hbtp]
        \centering
        \includegraphics[trim={0.0cm 0.0cm 0 0.0cm},clip,width=\columnwidth]{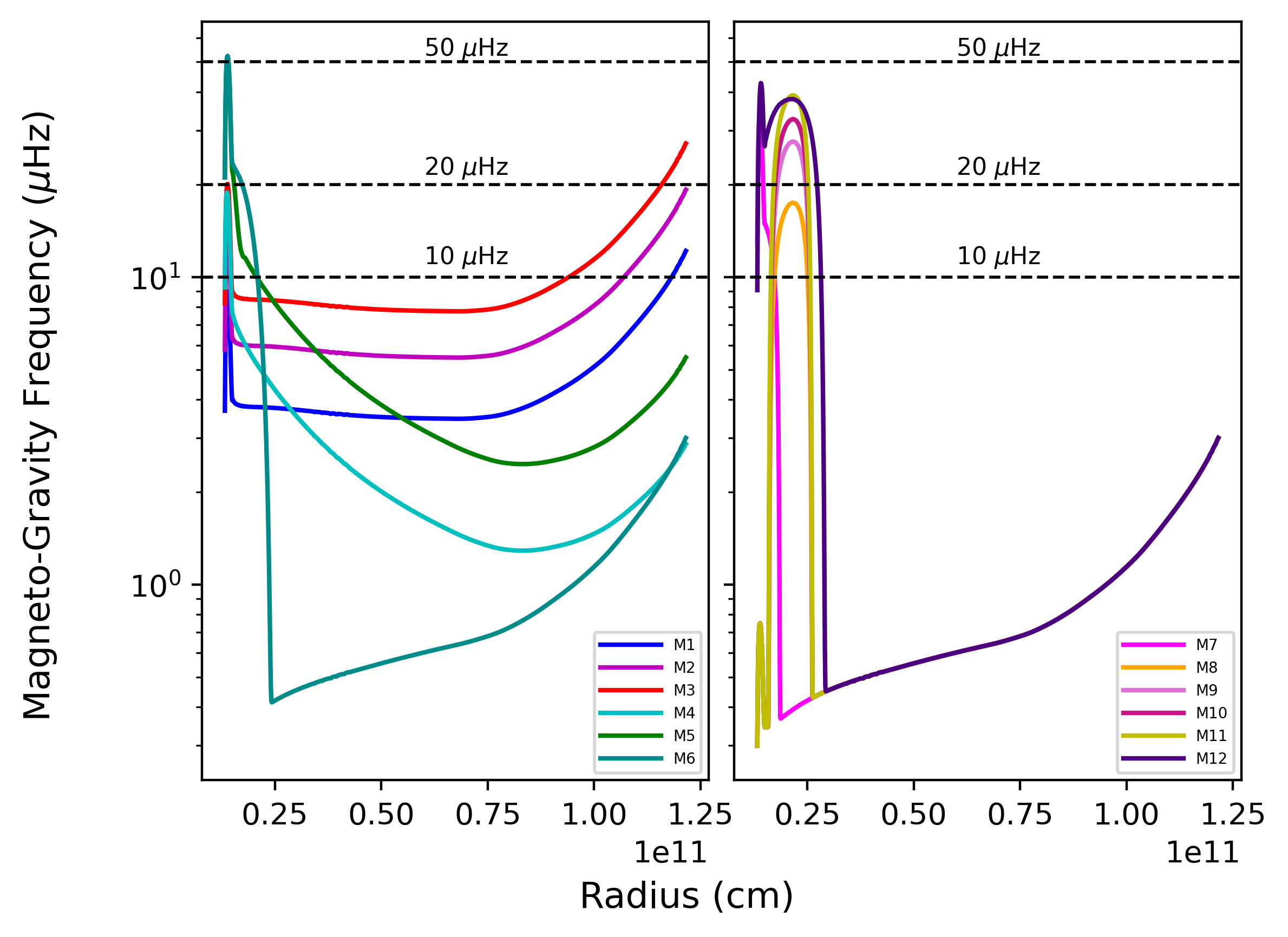}
        \caption{Magneto-gravity wave frequency calculated using Equation (1) for all magnetic models. Horizontal dashed lines represent frequencies as labeled. }\label{fig:magnetofreq}
        \centering
\end{figure}

In Fig.~\ref{fig:magnetofreq} we show the magneto-gravity wave frequency as a function of radius for our various magnetic models, calculated using Equation (1) (accounting for geometry).  There we see that, if we interpret the theoretical criteria as satisfied if {\it anywhere} in the propagation domain the frequency is below the magneto-buoyancy frequency, then {\it all} models, with the possible exception of M1, should have all modes below 10$\mu$Hz suppressed (this is also seen in Fig.~\ref{fig:initfields}). Similarly, using that theoretical criteria, models M5-M7 and M9-M12 would have all modes below 20$\mu$Hz completely suppressed for $m=1$.  We now turn to our results to see if our simulations are consistent with this picture.

\subsubsection{Global Fields}

\begin{figure*}[!hbtp]
        \centering
        \includegraphics[trim={0.0cm 0.0cm 0 0.0cm},clip,width=\textwidth]{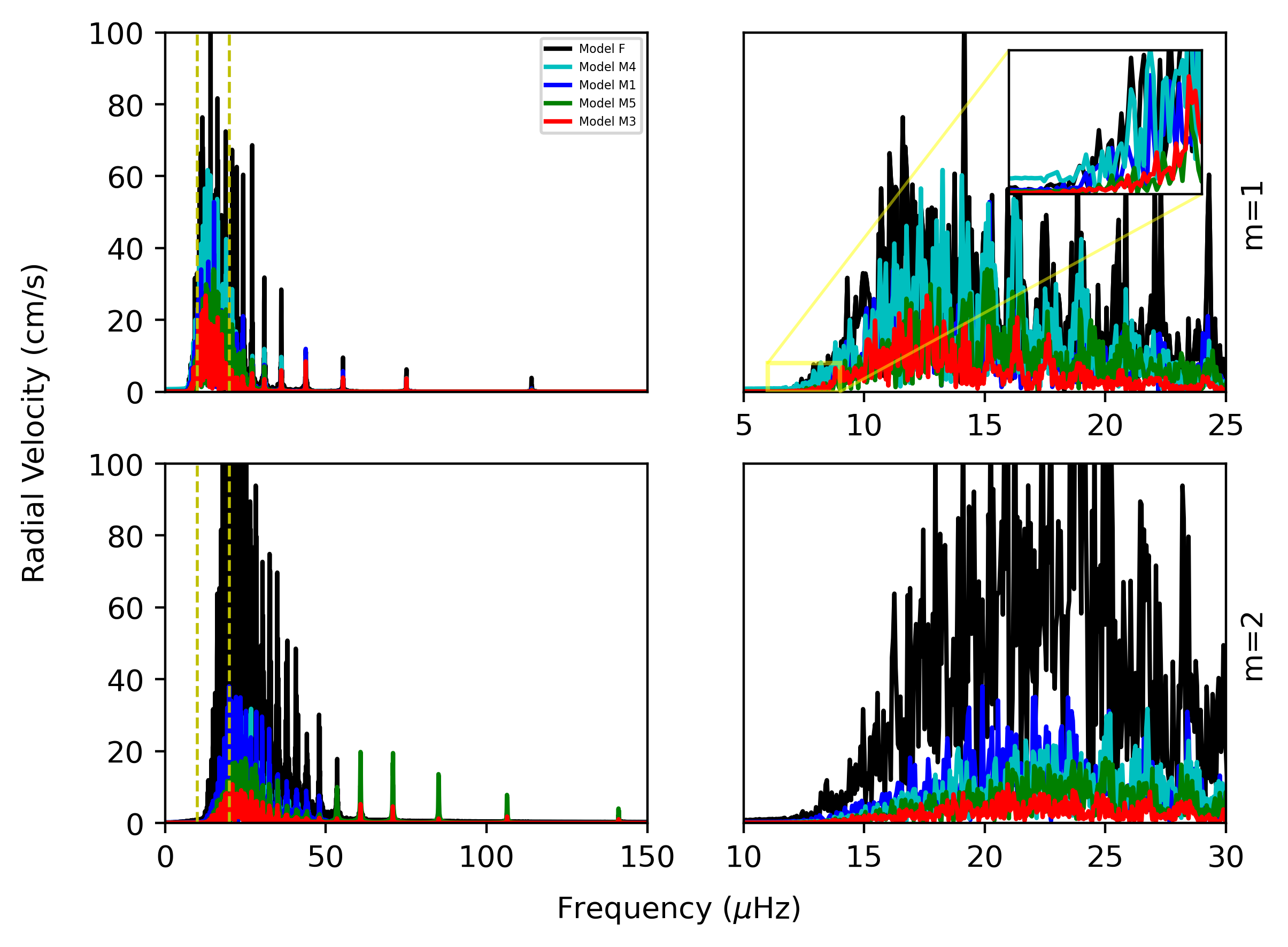}
        \caption{Radial velocity as a function of frequency and magnetic field strength for models M1,M3,M4 and M4 (as labeled) at both $m=1$ (top) and $m=2$ (bottom) at 85\% stellar radius. This shows the reduction in IGW amplitudes in the presence of a magnetic field and the dependence of that reduction on field strength.  The vertical yellow dashed lines represent 10$\mu$Hz, where M3,M4 and M5 would be completely suppressed and 20$\mu$Hz where model M5 would be completely suppressed according to Equation (1).  The inset shows a zoom of the lowest frequencies (6-9$\mu$Hz) indicating some {\it possible} suppression in a very narrow frequency range for the highest field strengths M3 and M5. } \label{fig:spec-field-amps}
        \centering
\end{figure*}

We show the radial velocity as a function of frequency for the global magnetic models M1,M3, M4 and M5 (M2 is very similar to M3 and M6 is very similar to M5, so we leave them out for clarity) in Fig.~\ref{fig:spec-field-amps} (with their hydrodynamic counterpart for comparison). There we see that the magnetic field has a clear impact on the amplitudes of the modes.  Gravity mode amplitudes are reduced at nearly all frequencies, with lower frequencies generally reduced more than higher frequencies. At $m=1$ amplitude reductions between 10-20$\mu$Hz, where we would expect modes to be suppressed based on Equation (2) only show an amplitude reduction of $\sim$4 in the strongest field cases M3 and M5.  However, within the inset of Fig.\ref{fig:spec-field-amps} we see possible suppression at lower frequencies, $\sim$8$\mu$Hz.  This will be revisited in Sec.\ref{sec:quantcorrect}.

At $m=2$ we see a larger reduction in mode amplitude at low frequency.  This is somewhat expected as the theoretical critical field strength (Equation 2) is reduced by a factor of $m$.  Higher frequency modes are reduced little in most cases.  For M5 the amplitudes are actually $\it{higher}$ than their hydrodynamic counterpart, indicating that a strong magnetic field can facilitate energy transfer to higher frequencies, rather than just dissipating the energy.  Overall, we are seeing some consistency with the theoretical prediction of mode suppression, but we clearly lack theoretical suppression at the expected frequencies.

\subsubsection{Radially confined fields}

Models M1-M6 represent field configurations likely for a main sequence star, where the magnetic field threads the entire star but is much larger within the convection zone and CRB than the radiative region.  For Red Giants, \cite{fuller2015} originally considered a strong, highly localized radial field.  The criteria described above and developed by \cite{fuller2015} and \cite{lecoanet2016} do not require a highly localized field, only that the radial field strength reach that criteria somewhere in the domain. However, it is worth investigating whether a localized field could impact the suppression of g-modes.  Models M7-M12 represent such highly localized fields.  In these highly localized models we are able to run both larger field strengths and lower magnetic diffusivities, since the field does not interact (much) with the convection.  However, we note that when imposing a strong, highly localized radial field, a strong, highly localized toroidal field,  will develop in order to satisfy the divergence-free condition. While this field could be strong, its interaction with predominantly horizontal g-modes is still weaker than with the vertical field. \cite{durfeldt26} estimated that a horizontal field would have to be a hundred times larger than the radial field to have a significant impact, within their linear approximation.

\begin{figure}[!hbtp]
        \centering
        \includegraphics[trim={0.0cm 0.0cm 0 0.0cm},clip,width=\columnwidth]{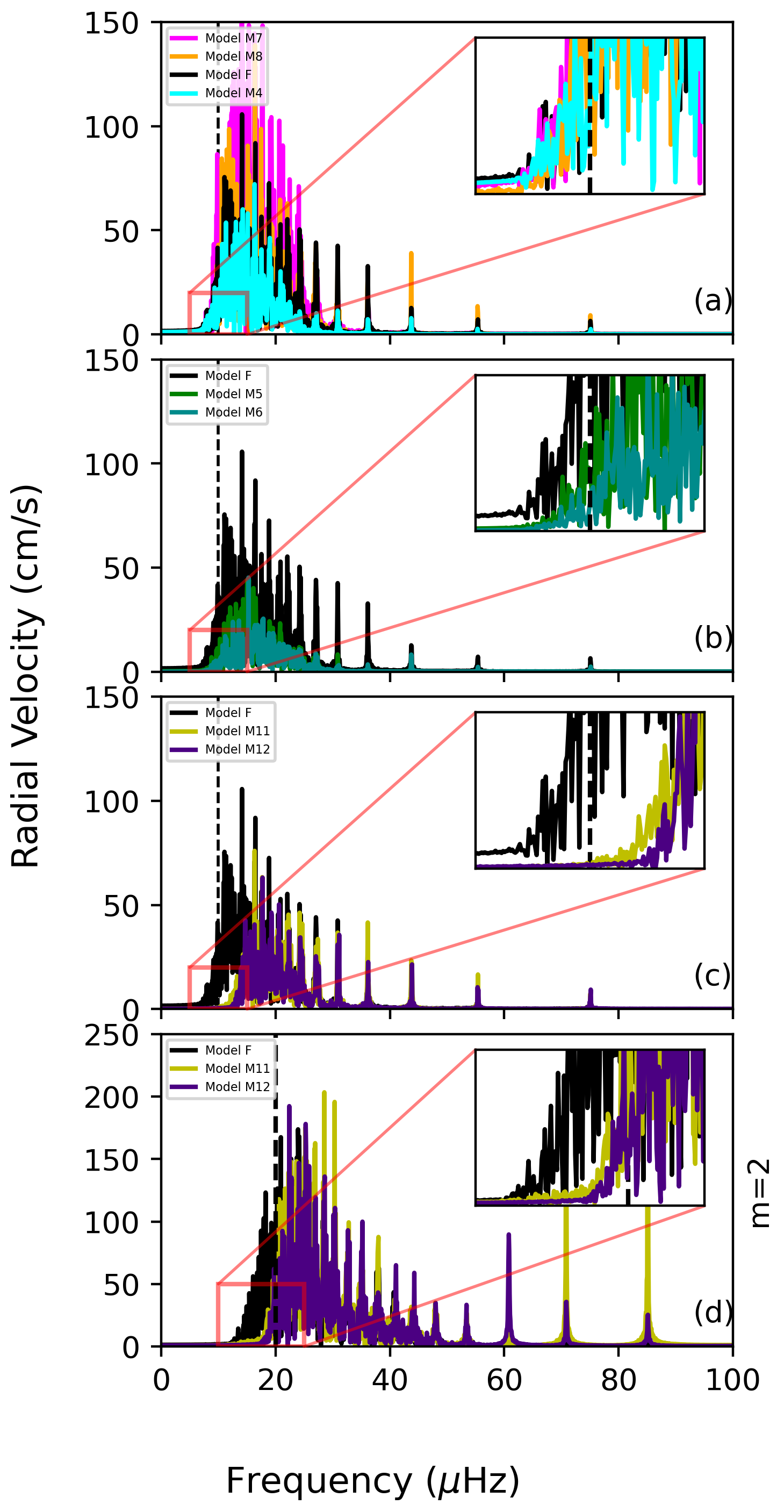}
        \caption{Amplitude comparison for various local and global models.  All spectra are shown near the top of the domain at 0.85R*.  Insets are zoomed between 5-15$\mu$Hz for (a)-(c) and 10-25$\mu$Hz for (d).  Vertical dashed lines show 10$\mu$Hz for (a)-(c) and 20$\mu$Hz for (d). (a) Shows a comparison of confined and global models with the same peak field strength (M4,M7 and M8). (b) Shows comparison of models which both span the convection zone, but have different strength and structure within the radiative region (M5 and M6). (c) Shows a comparison of local models, both at high field strengths but spanning slightly different radial extents (M11 and M12). (d) Shows the same models as in (c), but for $m=2$.}\label{fig:localfields}
        \centering
\end{figure}

Fig.~\ref{fig:localfields} shows four different comparisons of various local and global models.  In Fig.~\ref{fig:localfields}a we show Models M4, M7 and M8 alongside the fiducial hydrodynamic counterpart (always in black). Models M4, M7 and M8 all have the same peak magnetic field strength of $2\times 10^{5}$G, M4 has that peak within the convection zone and at the CRB, M7 has that peak within the \bvf{} spike (radiation zone) and M8 has that peak just outside the \bvf{} spike.  We can see that the localization of the field has no impact on the suppression of modes - none of the modes are suppressed in the local or global case.  From Fig.~\ref{fig:initfields} and Fig.~\ref{fig:magnetofreq} one can see that all of these models should have waves below 10$\mu$Hz suppressed, if the criteria in (1) were accurate.  However, one can clearly see that Model M4 has nearly all amplitudes reduced (though not suppressed), while both M7 and M8 show an {\it increase} in mode amplitudes, with M8 showing a significant increase across nearly all frequencies and particularly at low frequencies.  The reduction in amplitude for M4 is consistent with the reduction of driving due to reduced rms velocities as seen in Fig.~\ref{fig:vrms}.  The increase in g-mode amplitudes in M7 and M8 is due to the fact that the field itself can act as a source of waves (see Section.~\ref{sec:magspec}.) 

Next, in Fig.~\ref{fig:localfields}b, we compare models M5 and M6. Both of these models have a field strength  of $5\times 10^{5}$G within the convection zone and across the \bvf{} spike, but have very different amplitudes and radial structure in the rest of the radiative region. Model M6 has a higher field strength just outside the \bvf{} spike which extends a little further into the radiation zone than M5. In the spectrum, we see only a very small difference where M6 has slightly lower amplitudes at the lowest frequencies, consistent with its slightly higher field strength just outside the convection zone.  Again, we see amplitude reduction across all frequencies.  There is possible suppression, again around 8-10$\mu$Hz, but no suppression above that, despite the criteria indicating that modes below 20$\mu$Hz should be completely suppressed for both these models, as seen in Fig.~\ref{fig:magnetofreq}.

We show models F, M11 and M12 in Fig.~\ref{fig:localfields}c.  Models M11 and M12 both have a peak field strength of 1MG.  There we clearly see suppression of modes below $\sim 12\mu$Hz. Model M12, which spans a slightly wider range within the radiation zone, shows suppression shifted very slightly to higher frequencies, indicating that the radial extent of the strong field layer may play some (small) role in the frequency of suppression (as also seen between Models M5 and M6). 

Finally, we show models F, M11 and M12 for $m=2$ in Fig.~\ref{fig:localfields}d.  There we see that suppression is seen out to approximately 20$\mu$Hz for M11 and M12.   This demonstrates that the scaling with horizontal wavenumber is qualitatively consistent with \cite{fuller2015}.

\begin{figure*}[!hbtp]
        \centering
        \includegraphics[trim={0.0cm 0.0cm 0 0.0cm},clip,width=\textwidth]{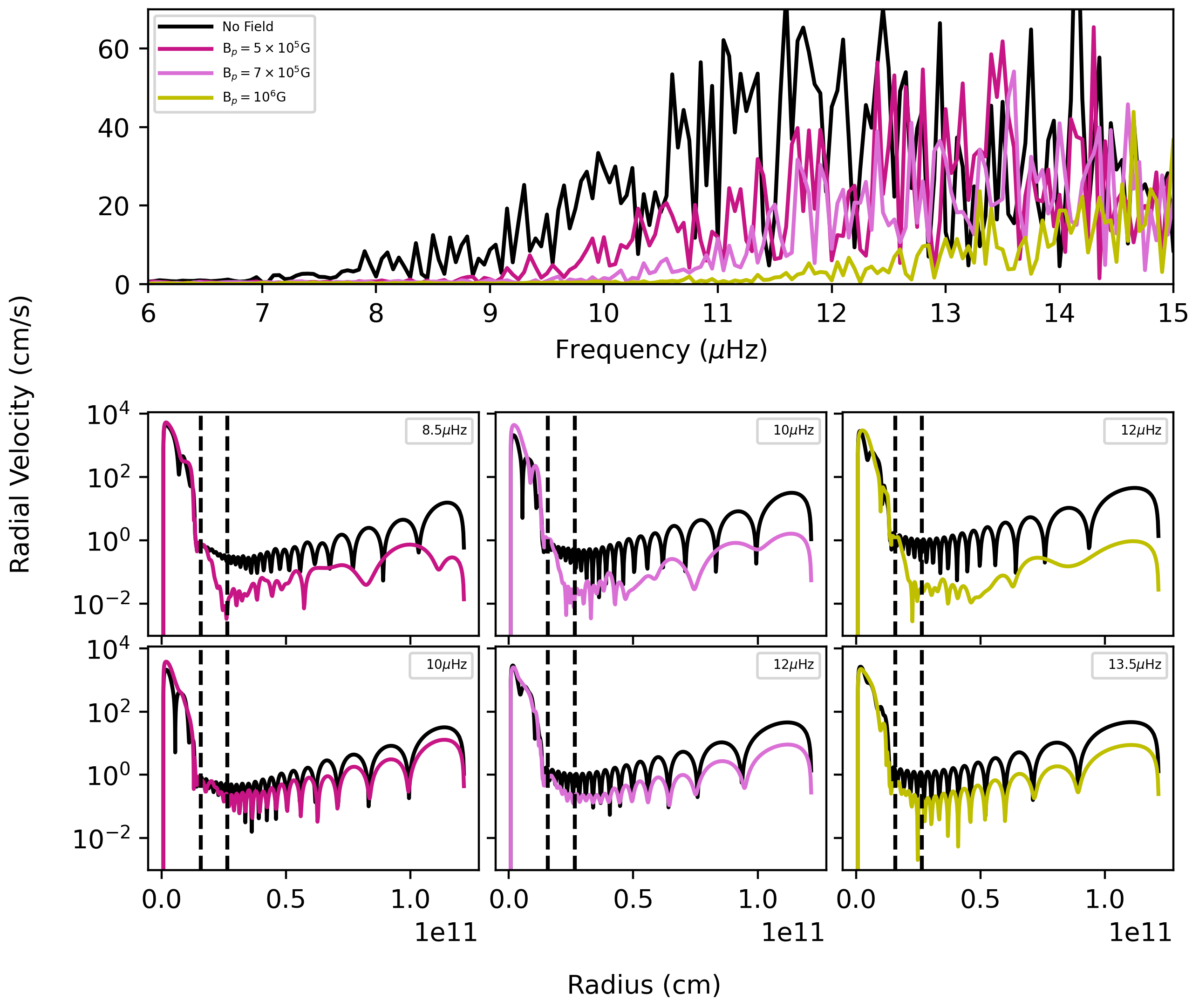}
        \caption{Suppression of g-modes.  The top panel shows a zoom in of the low frequency spectrum of local models with varying peak field strength, as indicated.  Showing that between 5$\times 10^{5}$G and $10^{6}$G low frequencies are progressively suppressed.  The middle row shows the radial structure of each of these models at the indicated frequency, which is the largest frequency at which the modes are suppressed by the "strong" criteria.  Bottom row shows slightly higher frequencies where modes are suppressed by the "weak" criteria, but not the "strong" criteria.}\label{fig:radstruct}
        \centering
\end{figure*}

We do not see any magnetic frequency shifts in any of these spectra.  This is not too surprising given the magnetic field strengths within the radiative regions of the simulations and the width of these stochastically driven modes. We show in Appendix \ref{app:mag} by deriving asymptotic magnetic frequency shifts in this 2D equatorial framework that the shift would be of the order of $10^{-2} \, \mu Hz$ at $20 \, \mu \rm Hz$ and $10^{-3} \, \mu \rm Hz$ at $40 \, \mu \rm Hz$. While we have sufficient frequency resolution to detect such a shift (at 20$\mu$Hz), the mode widths due to stochastic forcing and thermal diffusion are larger than the shifts.  Therefore, no shift is detected. 

Still, this demonstrates that a magnetic field has an impact on the amplitudes of the g-modes before it has an (observable) impact on their frequencies, at least for these stochastically generated modes.  The impact on coherently forced modes will be investigated in a separate paper. 

\subsubsection{A Quantitative Correction} \label{sec:quantcorrect}

Since we are seeing suppression, but at a much higher field strength than expected, or equivalently, at lower frequencies than expected, it is now worth determining if Equation (2) can be used, but with a correction to account for non-linear effects.  As can be seen in Fig.~\ref{fig:initfields}, the critical field strength using Equation (2) for a 10$\mu$Hz wave varies significantly with radius between a few $\times 10^{3}$G near the surface of the simulated domain to a peak around $10^{5}$G at the CRB, with a local minimum at the \bvf{} spike.  For this comparison we will use only models M9-M11 and standardize the critical field strength from Equation (2) to be that at the location of the peak field strength of those models. 

As can be seen in Fig.~\ref{fig:localfields} we achieve suppression at 1MG, possible suppression at 5$\times 10^{5}$G, and no suppression at 2$\times 10^{5}$G. In Fig.~\ref{fig:radstruct} we compare models M9,M10 and M11, showing a zoom of these models around the region of suppression (top panel). We use these three models as they have the exact same radial structure and position of their peak field strength.  The only difference is the amplitude of their peak field strength.  In Fig.\ref{fig:radstruct} we see that, as the field strength increases, suppression occurs out to higher frequencies, dependent on the imposed field strength.  The model with $B_p=5\times 10^{5}$G shows suppression out to $\sim 8-9 \mu$Hz, while the model with $B_p=7\times 10^{5}$G shows suppression out to $\sim 10 \mu$Hz and a 1MG field shows suppression out to $\sim$12$\mu$Hz. 

To determine the maximum frequency at which each model shows suppression we look at the radial structure of the low frequency modes.  We show the radial structure of the highest frequency which satisfies the "strong" suppression criteria in the middle panel of Fig.~\ref{fig:radstruct}.  This is the frequency where the mode's amplitude is reduced by at least an order of magnitude {\it and} it loses its gravity wave structure, which can clearly be seen.  We find that the suppression roughly follows the $\omega^2$ pattern predicted by \cite{fuller2015} and \cite{lecoanet2016}.  If we take our peak field strengths in those three models and divide them by the Fuller critical field strength from Equation (2) at the radii of those peaks, we find a factor of 10.3, 9.5 and 9.6, respectively.  We note that if we used the minimum critical field in the domain that this ratio would be significantly higher, hence this factor 10 pre-factor is a lower limit correction.

Given this we can infer that our nonlinear MHD simulations recover the critical field criteria developed by \cite{fuller2015}, but require a prefactor of approximately 10. Though this is only three data points, we note that we recover the same suppression frequency of 8.5$\mu$Hz for Models M5 and M6, which have the same peak field strength as M9, and 12$\mu$Hz for Model M12, having the same peak strength as M11, adding a few data points qualitatively. Unfortunately, the simulations are limited to this small range of field strengths as, for field strengths higher than this we start to see magnetic instabilities, which we do not attempt to resolve, and at lower field strengths, we see no suppression. While we do see complete suppression at $m=2$, it is far less consistent with the theory and more dependent on the exact field geometry.
\begin{figure*}[!t]
        \centering
        \includegraphics[trim={0.0cm 0.0cm 0 0.0cm},clip,width=\textwidth]{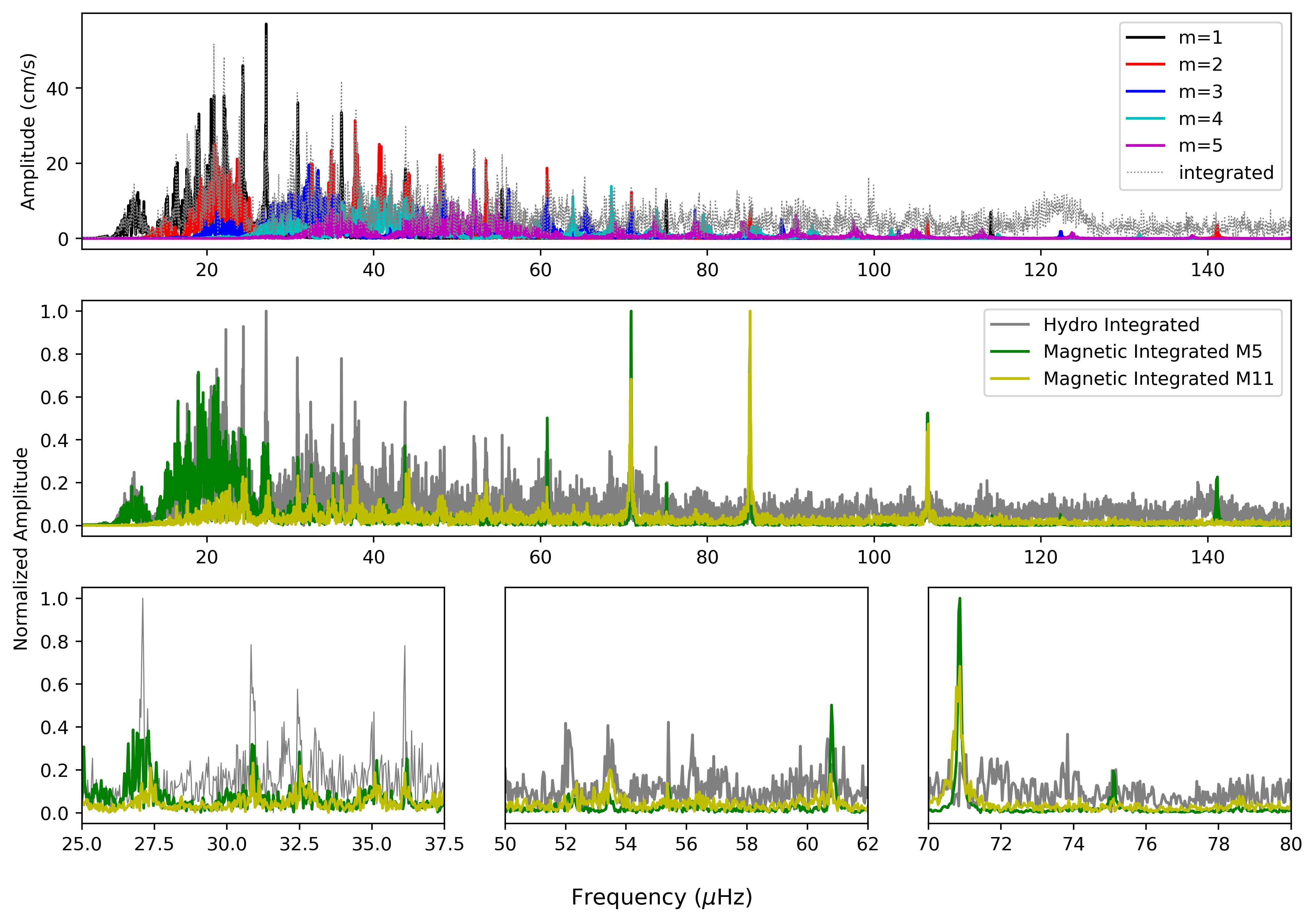}
        \caption{(a) Real space frequency spectra for Model F (grey) alongside spectra for given horizontal wavenumbers (colors as indicated in the caption). (b) Normalized real space spectra for Model F (grey) and Model M5 (green) and Model M11 (yellow), (c)-(e) Zoom in around various modes (normalized). The addition of a magnetic field clearly reduces the noise level of the spectra.  One can also see that both magnetic models show enhancements of the $m=2$ modes at higher frequencies, but particularly M11.}\label{fig:realspacespec}
        \centering
        
\end{figure*}

In the bottom row of Fig.\ref{fig:radstruct} we plot the radial wave structure at frequencies just {\it slightly} larger than their "suppression" frequency. There we see that the amplitudes are reduced, but that the modes retain their g-mode structure. We therefore conclude that the theoretical prescription requires a higher field strength at a given frequency to lead to mode suppression, under the "strong" criteria. However, reduction in amplitude ("weak" criteria) is gradual and happens at much lower field strengths.  Given this, it is now important to consider suppression more from an observational perspective. Until now we have focused on spectra at particular horizontal wavenumber, $m$, allowing us to directly identify standing modes and relate them to the dispersion relation and theoretical suppression criteria.  Of course, this is not what is seen observationally, so despite the artificial geometry, we now turn to real space spectra.

\subsection{Real Space Spectra}\label{sec:rspec}

To mimic a real space spectra, we pick a point near the top of our domain and take the fourier transform of that time series.  The resulting spectrum has contributions from all horizontal wavenumbers, which is more akin to frequency spectra obtained from observed photometry.  The horizontal position we choose makes no difference to the overall results/spectrum.  However, the radial position becomes increasingly noisy the closer to the simulated surface we go.  To avoid this we choose a radius  at 75\% of the simulated radius, corresponding to just below the inflection point of the \bvf{} frequency profile seen in Fig.~\ref{fig:stellar_param}.  We show this for Model F in Fig.~\ref{fig:realspacespec} where we show the grid space frequency spectrum, alongside the frequency spectra for $m$=1--5, so that we can identify the peaks in the real space spectrum with individual wavenumbers (top).  We see the dominant peaks are from m=1,2 and occasionally 3.  We then compare the (normalized) hydrodynamic real space data with the (normalized) magnetic real space data in the middle panel for model M5 (green) and M11 (yellow), with zoom ins around individual lines in the bottom row.  Similar to what is seen in Fig.~\ref{fig:spec-field-amps} the magnetic case has relatively more energy at higher frequencies and at $m=2$. 

If we interpret this spectra as something more akin to an observational spectra, Model M11 may be interpreted as having suppression out to $\sim$ 12$\mu$Hz - most of the low frequency $m=1$ modes are gone, particularly compared to $m=2$ modes. On the other hand, M5 would not look suppressed, the low frequency $m=1$ modes are still present, compared to $m=2$.  Yet M5 has a field strength which is also, theoretically, super-critical. 


Observationally, it is the "weak" definition that is relevant. Studies tackling mode suppression in intermediate-mass stars have considered suppressed modes as ones not meeting some signal-over-noise threshold along the iterative pre-withening \citep{lecoanet22, durfeldt26}. This means that even though a mode would not be completely suppressed following the "strong" definition, it could be considered so if its amplitude was not sufficiently dominating the noise level. On the RGB, mode suppression was inferred by a loss of visibility of the dipolar modes compared to the radial ones, which are not directly impacted by magnetism \citep{stello2016}. It might also be that modes would not be "strongly" suppressed but "weakly", where the noise level would be high enough to hinder the visibility of dipolar modes. In both cases, the inference of a critical field strength on observational data based on a theoretical criterion seems to suffer from the observed frequency difference between the "strong" and "weak" suppression seen in our work.

Indeed, if one were to use the M11 spectrum to infer an upper limit on the field strength based on the frequencies below $\sim$15$\mu$Hz being suppressed, using the linear criteria they would an infer an {\it upper} limit of approximately 10$^{5}$G (if they considered a near core radial field), yet the actual field strength is 10$^{6}$G. Similarly, one might infer that the M5 spectra did not have a significant near core field, since no frequencies are suppressed, yet it has a near core field strength of 5$\times 10^{5}$G.

Beyond mode reduction/suppression, we see a clear difference in the noise level between the magnetic and non-magnetic cases.  The reduction in noise around each peak also reduces the total full-with at half maximum (FWHM) of the peak, therefore leading to an increase in the overall mode lifetime. We generally see a decrease in FWHM of a factor of two from non-magnetic to magnetic, corresponding a similar increase in mode lifetime. Further, the reduced noise in the magnetic models also corresponds to an increase in the signal-to-noise ratio (SNR) of a factor of $\sim$5.3 (M5) and $\sim$7.0 (M11) on average.

In addition to reducing the noise, when looking at the normalized spectra we also see that the dominant modes shift to higher frequencies and $m=2$ in the magnetic cases. The frequency with maximum amplitude in the (integrated) hydrodynamic spectrum occurs at 27$\mu$Hz (at this radial level), while for M5, it is at 61$\mu$Hz and for M6, it is at 71$\mu$Hz as seen in Fig.\ref{fig:realspacespec}.  One can see that overall, in the magnetic case, energy is transferred to higher frequencies (lower radial orders at the same $m$) and higher $m$.

\begin{figure}[!hbtp]
        \centering
        \includegraphics[trim={0.0cm 0.0cm 0 0.0cm},clip,width=\columnwidth]{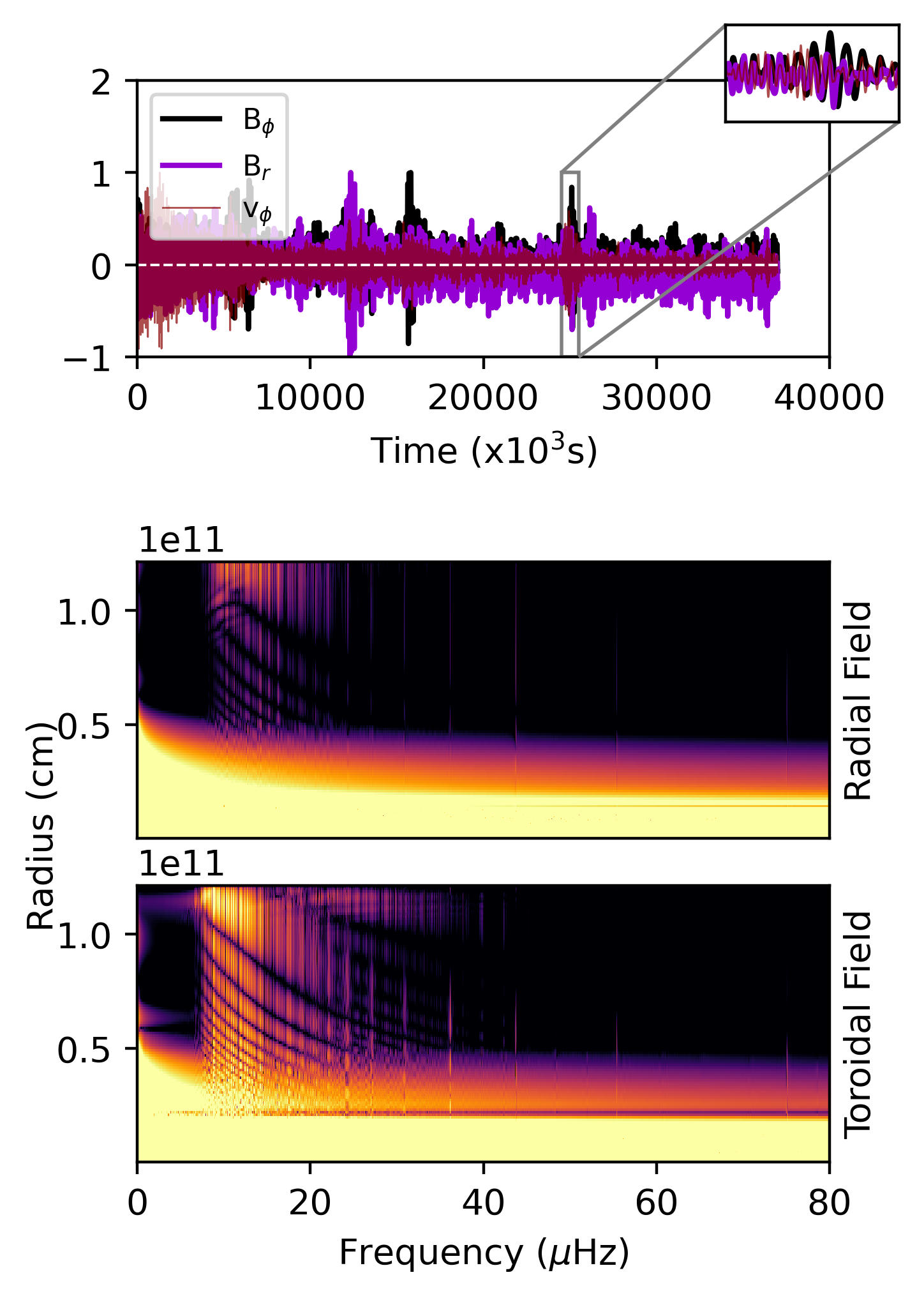}
        \caption{Normalized time series of $B_{\phi}, B_r$ and $v_{\phi}$ for M3 at 0.82R$_{*}$(top) for $m$=2 (since $m$=1 has no time variation), clearly showing varying frequencies and a decaying amplitude of the velocity (IGW).  Frequency spectra are shown in the middle (radial field) and bottom (toroidal field), showing that the magnetic field oscillates at the same frequencies of IGW, in addition to lower frequencies defined by the local Alfv\'en frequency. \label{fig:bfieldosc}}
        \centering
\end{figure}

\subsection{Magnetic Impact on the Evanescent Tail}
\begin{figure}[h]
        \centering
        \includegraphics[trim={0.0cm 0.0cm 0 0.0cm},clip,width=\columnwidth]{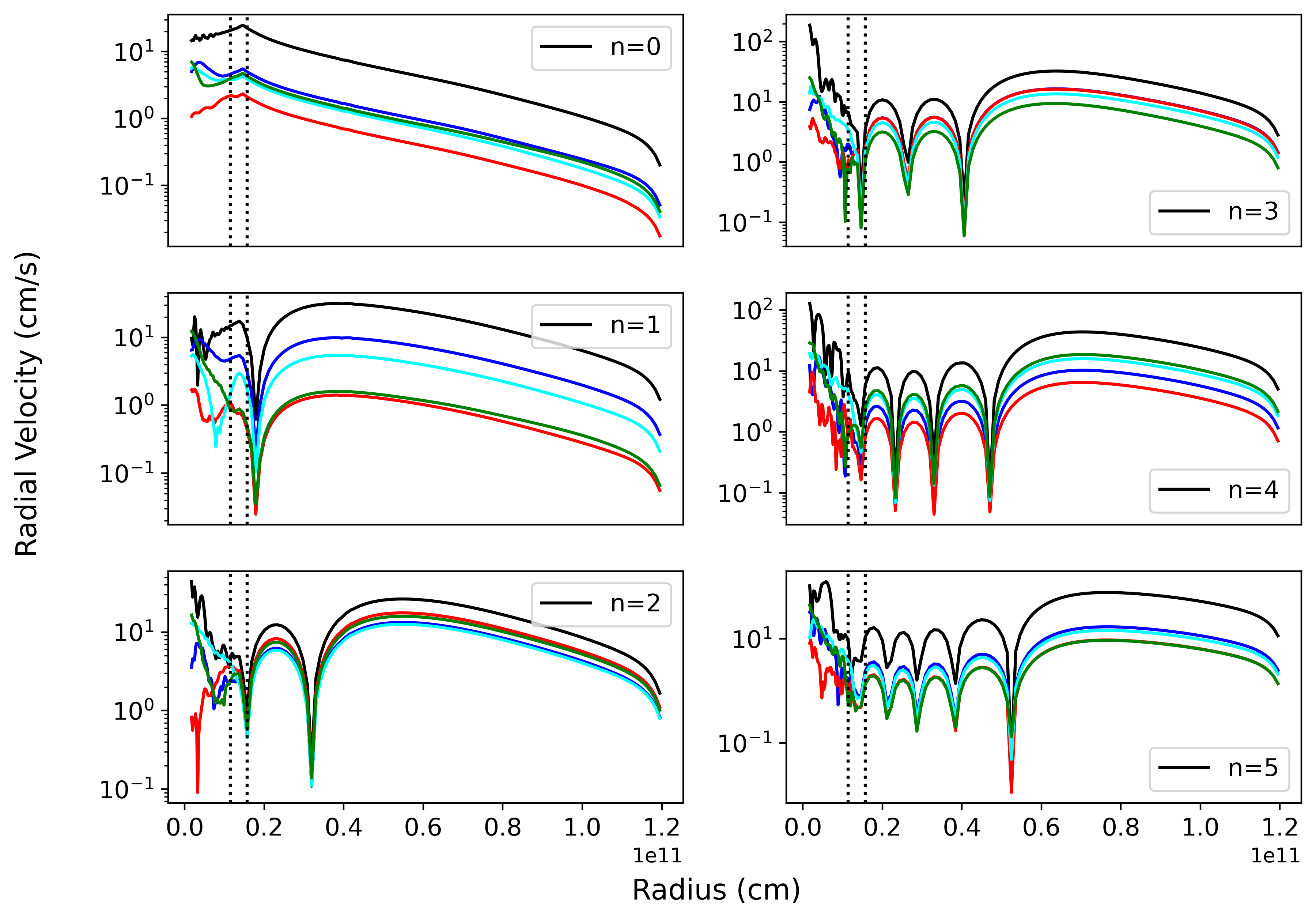}
        \caption{Radial Velocity at $m$=1 as a function of  radius for six different radial orders (as indicated in figure). Colors match those seen in Fig.~\ref{fig:spec-field-amps}.  The wave amplitude is reduced, both in the radiative envelope and in the convective core.  \label{fig:evanescent}}
        \centering
\end{figure}

Given the impact of the field on the g-modes in the radiation zone, it is worth investigating whether the field has an impact on the evanescent tail of the g-modes in the convection zone as it could have important implications in the Sun \citep{appourchaux2010,fossat2017,schunker18}.  We show the amplitudes of individual modes as a function of radius, including the convection zone, in Fig.~\ref{fig:evanescent} for our "global" simulations M1 and M3-M4.  There we see that at every radial order (frequency) the presence of the magnetic field reduces the amplitude of the "evanescent tail". We have put evanescent tail in quotes here as it is clear that the mode does not behave simply as an evanescent decay in the convection zone. Indeed the mode interacts with convection and its amplitude is increased in the convection zone, except at the lowest radial order.  Still, it remains that the amplitude of this frequency of motion is decreased in the magnetic cases, compared to its hydrodynamic counterpart, just as it is within the radiative envelope.  There is clearly some dependence on magnetic field strength and radial order, with lower radial orders and higher field strengths showing a larger effect.  Overall, at these frequencies there is an approximately order of magnitude reduction in the amplitude of the evanescent tail (which is larger than the reduction of rms velocities). Given that we expect a strong toroidal field in the solar tachocline and perhaps a weak radial field in its radiative region, if the trend is similar, this could have significant consequences for the hunt for g-modes in the Sun. 

\subsection{Magnetic Field Oscillations}\label{sec:magspec}

There are many factors that may contribute to the theoretical prediction in \cite{fuller2015} requiring quantitative refinement.  The two primary ones are that their work neglects the presence of a toroidal field and that it neglects the back reaction of the field on the waves.  Though \citet{durfeldt26} considered a toroidal field, they did so in only a passive manner. When considered consistently, with back reaction, the two reasons together likely explain the discrepancy.  When a wave interacts with a field, the field itself oscillates.  If that field is constrained to be just radial it is not able to generate a (significant) Maxwell stress.  However, if both a radial and toroidal field oscillate together, this can easily lead to a Maxwell Stress, which can generate waves.  Whether the inclusion of the field then leads to an enhancement of waves or a impediment to waves is a balance between the generation of the waves and the reduction.

Fig.~\ref{fig:bfieldosc} shows the time series of  of  $B_{\phi}, B_r$ and $v_{\phi}$ (top), alongside the frequency spectrum of both $B_{r}$ (middle) and $B_{\phi}$ (bottom) for Model M3 at $m=2$.  There, we can see in that global model that the field oscillates at the gravity wave frequencies and shows similar gravity wave structure.  Perhaps more importantly we see strong power in both radial and toroidal field at lower frequencies.  It is likely these fluctuations in field, which result in a Maxwell stress, that lead to an additional source of waves.  Of course in local models, the oscillations are (mostly) confined to the region of strong field, still the Maxwell stress generated there can produce gravity waves propagating outward (or inward as the case may be), as we see in Fig.\ref{fig:localfields}a.

\begin{figure}
    \centering
    \includegraphics[trim={0.0cm 0.0cm 0 0.0cm},clip,width=\columnwidth]{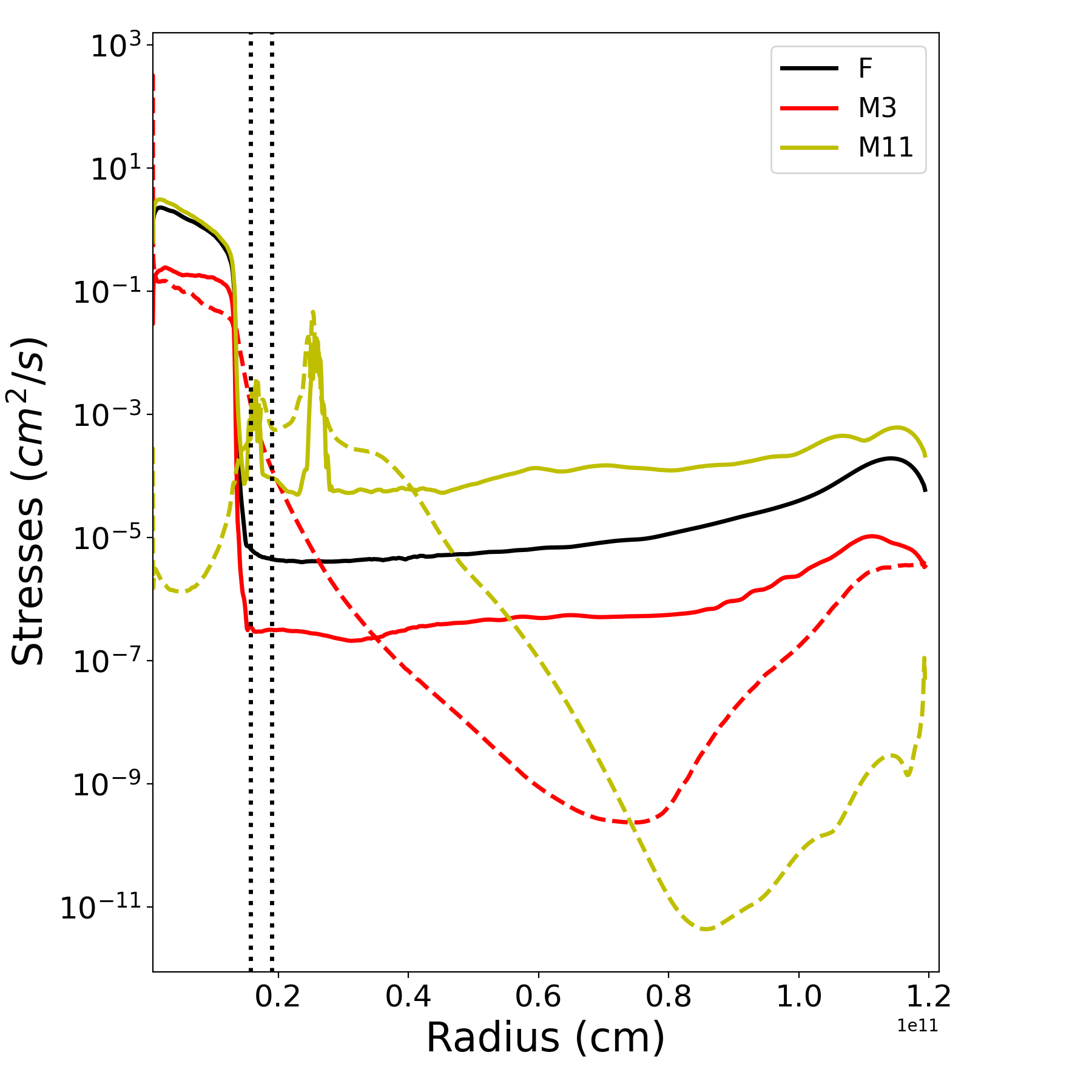}
    \caption{Time-averaged radial profiles of Reynolds and Maxwell stresses shown for models M3, and M11 throughout the star over 1 year of the simulation. Reynolds stresses are shown as solid lines, while Maxwell stresses are shown as dashed lines. The two vertical dotted black lines show the location and width of the spike in BV frequency. Spikes in the Reynolds and Maxwell stresses in M11 (local model) are due to the generation of currents at the sharp gradients in field strength, these currents then generate vorticity, leading to a local enhancement of both stress terms.}
    \label{fig:stresses}
\end{figure}

Another way to view this is by considering the source term for IGW.  For stochastically generated g-modes, the source term is the Reynolds stress \citep{press1981}. However, at the CRB magnetic fields are strong and hence also contribute through the Maxwell stress \citep{campos2010,jaupart2017,bessilla24}.  We show the Maxwell and Reynolds stresses in Fig.\ref{fig:stresses}, which demonstrates that not only are Maxwell stresses relevant at the CRB, but are indeed dominant.  This is true for all models.\footnote{Although we show only the longitudinal component of these stresses we have confirmed that these terms are dominant over radial terms and hence, are likely the main source of g-modes.}  This is likely the reason we do not see complete suppression until much larger field strengths and certainly explains the higher amplitude g-modes seen in some local models. None of the previous work on the suppression of g-modes by a magnetic field has considered the field as also a source of waves.

Though it is impossible in nonlinear simulations such as these to track exactly what stress is driving what waves, its clear that the Maxwell stress must also contribute to wave generation.  Moreover, given the natural frequencies of the field oscillations, as seen in Fig.~\ref{fig:bfieldosc} it is likely those stresses contribute precisely where waves might be suppressed. We note that the strong Maxwell stress is due to there also being a strong toroidal field in the region.  This toroidal field can be generated through nonlinear shear, or merely to satisfy the solenoidal constraint, either way its presence is not just likely, but required. Furthermore, the nonlinear effects its inclusion leads to are fundamentally important to this problem.

\section{Discussion}

Previous investigations of magnetic interactions with internal gravity waves have primarily employed linear theory or numerical calculations focused on individual waves. Although these studies have provided important insights into magnetic modifications of wave propagation, they have not addressed the combined effects of nonlinear wave–field interactions, magnetic field evolution, and wave excitation on the global oscillation spectrum. Our simulations allow us to investigate these effects and reveal three important results.

First, the reduction of g-mode amplitudes with increasing magnetic field strength is gradual, rather than occurring abruptly at a particular critical field strength. This arises both from the direct interaction between the magnetic field and the waves, previously investigated by \citet{loi2018} and \citet{muller26a}, and due to the influence of the magnetic field on convection and, consequently, wave excitation. The latter effect has not previously been quantified in this context. The strongest reduction occurs at low frequencies, whereas higher-frequency waves are relatively unaffected or, in some cases, amplified. Consequently, increasing magnetic field strength progressively reduces the low-frequency portion of the spectrum, shifting its peak amplitude towards higher frequencies.

Second, achieving theoretical mode suppression, defined here as a reduction in amplitude of at least an order of magnitude {\it and} a loss of the g-mode wave structure, requires magnetic field strengths at least an order of magnitude larger than those predicted by commonly used linear criteria \citep{fuller2014,lecoanet2016}. The distinction between reduced amplitudes and observationally undetectable modes necessarily depends on the sensitivity of the observations. Nevertheless, our results suggest that magnetic field strengths inferred from the presence or absence of low-frequency modes require substantial revision.

Third, we find that the magnetic field can itself be an important source of waves, providing a possible explanation for the higher suppression thresholds. Although magnetic wave generation has previously been considered theoretically \citep{bessilla24}, its consequences for mode suppression have not been explored. Perturbations to a magnetic field with radial and horizontal components generate Maxwell stresses that can excite additional waves. This horizontal field does not have to come from nonlinear induction. The radial confinement itself leads to a horizontal component of the field through the solenoidal constraint. These strong horizontal magnetic fields and consequent Maxwell stress fundamentally change the wave dynamics, producing enhanced wave amplitudes across much of the spectrum.  This is particularly evident in our local models, both because they lack the reduced amplitude effect on the convection and because their localization leads to a particularly strong, localized, horizontal field. We suggest that this additional excitation partially offsets magnetic attenuation, requiring stronger fields to achieve substantial suppression. However, the range of field strengths over which suppression can occur without the onset of magnetic instabilities appears limited. Our strongest-field and steepest-gradient models show indications of such instabilities, although a detailed investigation is beyond the scope of this work.

These findings have important implications for magnetic field strengths inferred from mode suppression. Previous studies have employed suppression criteria in two distinct ways. In red giants, \citet{stello2016} used the reduced visibility of dipole mixed modes to infer {\it lower} limits on the radial magnetic field strength, based on the critical-field criterion of \citet{fuller2015}.  Conversely, \citet{lecoanet22} and \citet{durfeldt26} derived {\it upper} limits for four intermediate-mass pulsators from the absence of g modes below particular frequencies. Similar upper limits were inferred using this criteria for white dwarfs \citep{rui25wd} and hot subdwarfs \citep{ruima26}. 

Although this linear approach establishes conditions under which magnetic fields can strongly modify wave propagation, the resulting attenuation of the global oscillation spectrum also depends on wave excitation, nonlinear energy redistribution, and dissipation. Our results indicate that significantly stronger fields are required to produce substantial mode suppression when these effects are included. Consequently, at the very least, both lower and upper limits inferred using linear suppression criteria need to be revised upwards by an order of magnitude. Overall, inferring magnetic field strengths with this criteria should be used with caution.
 
More recently, \citet{shen2023} suggested that stochastic low frequency variability (SLF) could be affected by magnetic fields, possibly causing a change in its low frequency spectrum.  Others have recently suggested that this could be used to argue against IGW as a source of this variability, as magnetic fields may suppress those waves.  Again, such applications are questionable when nonlinear effects are included.  Magnetic fields can lead to mode suppression, but can also lead to wave generation.  Which one dominates depends on a host of factors which are not captured by linear theory.

An important observational test would be to compare magnetic field strengths inferred from mode suppression with independent measurements based on magnetic frequency shifts, particularly in partially suppressed red giants. Such comparisons have already revealed potentially relevant discrepancies. \citet{deheuvels2023} reported a red giant (KIC 6975038) for which the magnetic field inferred from frequency shifts exceeded the critical field estimated from partial suppression by a factor of 1.7, with a similar discrepancy reported by \citet{villate2026}. However, these comparisons require caution: \citet{deheuvels2026} demonstrated that perturbative calculations can overestimate magnetic field strengths near the critical field, highlighting the need for non-perturbative treatments. Applying such calculations to partially suppressed red giants would provide a valuable test of the relationship between magnetic field strength and mode visibility. Similar comparisons could be pursued for intermediate-mass main-sequence stars by applying recently developed magnetic frequency-shift diagnostics \citep{takata26,ihallaine2026} to pulsators exhibiting evidence of mode suppression \citep{lecoanet22,durfeldt26}.

Of course, our geometry is significantly different than that of Red Giants, where waves are generated from above and propagate initially {\it inward} instead of outward.  Moreover, the observations in Red Giants are of mixed modes, requiring coupling to the p-mode component.  Still, this change in geometry does not change the fundamental physics described here.  Standing modes must propagate both inward and outward to set up the standing mode, therefore they feel the global field the same as if they were (initially) propagating outward.  In either case, a magnetic field can indeed suppress g-modes, but it can also act as a source of g-modes.  There is no reason to expect the hypothesized magnetic layer in Red Giants would behave differently in this regard, though of course, the magnitude of the effect will require further investigation.

Our simulations extend previous linear treatments of wave–field interactions by capturing the nonlinear coupling between magnetic fields, convection, differential rotation, and gravity waves. This allows us to investigate not only the direct influence of magnetic fields on wave propagation, but also their effects on wave excitation and the redistribution of energy between frequencies and spatial scales, including global modes that lie outside the validity of the JWKB approximation. Although restricted to two dimensions, the simulations retain several key features of the stellar problem, including the interaction of predominantly horizontal wave motions with a predominantly radial magnetic field, realistic stellar stratification, and a self-consistently generated spectrum of waves. The two-dimensional geometry prevents the development of a self-sustaining dynamo, requiring the background magnetic field to be imposed and maintained. However, this also provides an important advantage: the field strength and geometry can be varied independently, allowing their effects on the global wave spectrum to be isolated.  Thus, while future three-dimensional calculations and studies of coherently excited g-modes will provide valuable extensions, the present simulations establish a more complete framework for understanding how dynamically evolving wave–field interactions modify the global oscillation spectrum of magnetised stars.

\vspace{0.2cm}
{\it Acknowledgements:}
We acknowledge support from STFC grants ST/X508615/1, UKRI1185 and ST/W001020/1.   L. Barrault acknowledges the support of the Austrian Academy of Sciences through the Doctoral Fellowship Programme (DOC) of the Austrian Academy of Sciences 27648. We thank Dominic Bowman and Saskia Hekker for extensive input on the observational context of this work.  ChatGPT (OpenAI) was used to assist with editing, LaTeX formatting, and refinement of plotting code. All scientific analyses, interpretations, and conclusions were made by the authors.


\vspace{5mm}





\bibliography{2Ms2DSpec}{}
\bibliographystyle{aasjournal}



\appendix 

\section{Expected magnetic shifts in a 2D equatorial framework}
\label{app:mag}
We first define the Lagrangian displacement vector as $\boldsymbol{\xi}(\boldsymbol{r},t) = \Re[(\xi_{r}(r)\boldsymbol{a}_{r}+im\xi_{\phi}(r)\boldsymbol{a}_{\phi})e^{im\phi-i\omega t}]$ with $(\boldsymbol{a}_{r}, \boldsymbol{a}_{\phi})$ the polar basis vectors. $\boldsymbol{b}$ is the magnetic perturbation, and $\boldsymbol{B}$ the background B field.
The Lorentz operator reads, not accounting for compressibility in this anelastic framework \citep{mathisbugnet21}:
\begin{equation}
    \delta\boldsymbol{\mathcal{F}}_{L} = \frac{1}{\mu_0 }[(\boldsymbol{\nabla}\times \boldsymbol{b})\times \boldsymbol{B_{0}} + (\boldsymbol{\nabla} \times \boldsymbol{B_{0}})\times \boldsymbol b] 
\end{equation}
From which the magnetic frequency shift reads:
\begin{equation}
    \delta \omega_{\rm mag} = -\frac{<{\boldsymbol{\xi}},\delta\boldsymbol{\mathcal{F}}_{L}/\rho>}{2\omega<\boldsymbol{\xi},\boldsymbol{\xi}>}
\end{equation}
$\omega$ being the unperturbed frequency. The full expansion of this shift contains numerous terms, like in the 3D case. Defining the cross term $\mathcal{C} = \xi_{r}B_{0,\phi}-\xi_{\phi}B_{0,r}$, we obtain:
\begin{align}
    \boldsymbol{\xi}^{*}\cdot[(\boldsymbol{\nabla}\times \boldsymbol{b})\times \boldsymbol{B_{0}}] = \left(\xi_{r}^{*}B_{0,\phi}-\xi_{\phi}^*B_{0,r}\right)\left(\frac{1}{r}\partial_{r}r\partial_{r}\mathcal{C}-\frac{m^{2}}{r^{2}}\mathcal{C}\right)
\end{align}
and
\begin{align}
    \boldsymbol{\xi}^{*}\cdot[(\boldsymbol{\nabla}\times \boldsymbol{B_{0}})\times \boldsymbol{b}] = \left(\xi_{r}^{*}\partial_{r}\mathcal{C}-\frac{im}{r}\xi_{\phi}^*\mathcal{C}\right)\left(\frac{1}{r}\partial_{r}rB_{0,\phi}-\frac{1}{r}\partial_{\phi}B_{0,\phi}\right)
\end{align}
Unlike in the high radial order case for which a term dominates in this expansion, the low radial orders of modes studied in this work forbids to draw a conclusion on the shifts produced between the second and the first part of the Lorentz operator, as the magnetic field can vary on shorter length scales than the local wavelength of the mode. However, our aim being only to explain the absence of measured frequency shift in our simulation, we restrict our argument to the lowest frequency modes (highest radial orders), for which this approximation is better supported.
A JWKB treatment holds \citep[][adapted to a 2D framework]{unno1989}:
\begin{align}
    \xi_{r} & = A_{W}\frac{|m|}{r^{3/2}\omega^{2}\rho^{1/2}}\frac{1}{\sqrt{k_{r}}}\cos\left (\int_{r_{a}}^{r}k_{r} dr -\frac{\pi}{4}\right) \\
    \xi_{\phi} & = -A_{W}\frac{N}{r^{3/2}\omega\rho^{1/2}}\frac{1}{\sqrt{k_{r}}}\sin\left (\int_{r_{a}}^{r}k_{r} dr -\frac{\pi}{4}\right)
\end{align}
Assuming that $\omega\ll N$ deep inside the radiative zone, with $k_{r}^{2} = \dfrac{N^{2}m^{2}}{r^{2}\omega^{2}}$ the large wavenumber of g-modes. $A_{W}$ is a wave-displacement amplitude that is unknown and of no relevance for the discussion.
Qualitatively, one can see that $\dfrac{\xi_{\phi}}{\xi_{r}}\propto \dfrac{N}{\omega |m|}$, as one can expect from g-modes.

\begin{figure}
    \centering
    \includegraphics[width=0.5\linewidth]{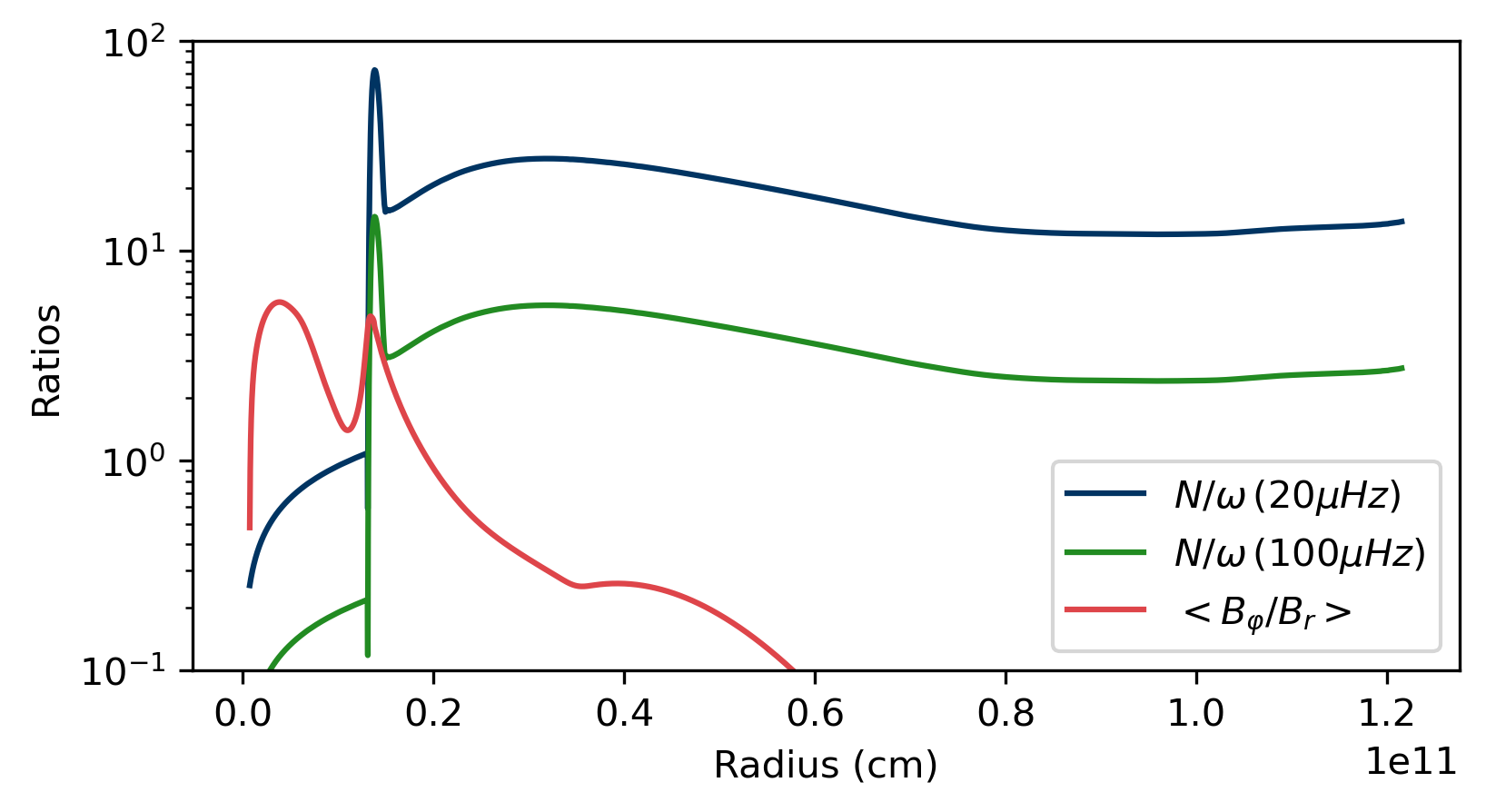}
    \caption{Ratios of the Brunt-Väisäla and the mode frequencies for the lower ($20 \, \mu Hz$, blue) and higher ($100 \, \mu \rm Hz$, green) frequencies, and ratio between the quasi-steady amplitudes of the azimuthal and radial components of the magnetic field.}
    \label{fig:ratios}
\end{figure}

\begin{figure}
    \centering
    \includegraphics[width=0.5\linewidth]{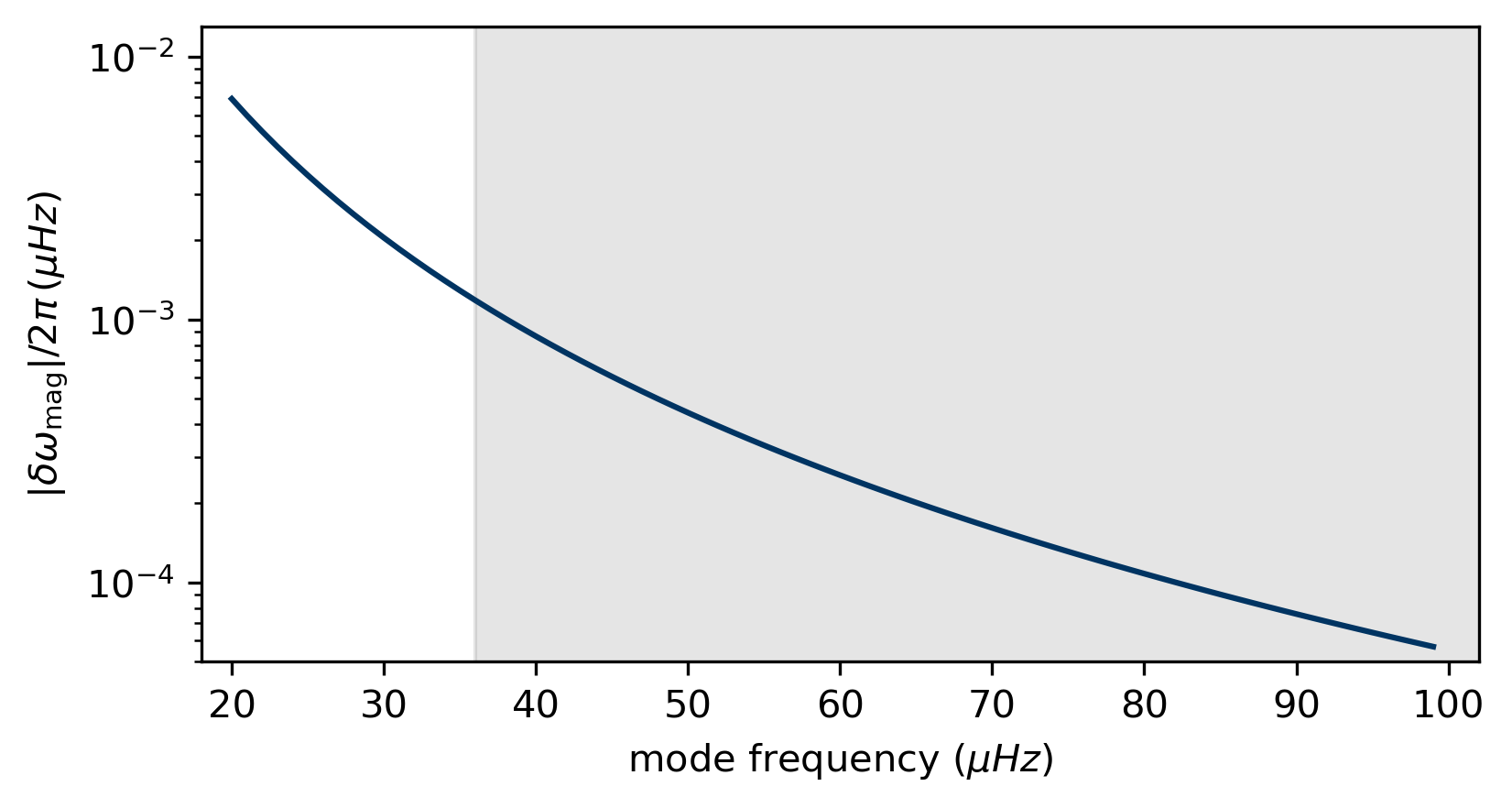}
    \caption{Asymptotic (linear) frequency shift obtained from Eq.\eqref{eq:shift} in term of the mode frequency, for the magnetic field profile of model M4. The grayed zone corresponds to modes with radial orders $n\le 8$ and indicates a non-reliability of an asymptotic analysis.}
    \label{fig:asym_shift}
\end{figure}

In this context, dominant terms in then Lorentz operator are:
\begin{equation}
    \boldsymbol{\xi}^{*}\cdot[(\boldsymbol{\nabla}\times \boldsymbol{b})\times \boldsymbol{B_{0}}] \approx -k_{r}^{2}|\xi_{\phi}|^{2}B_{0,r}^{2}-k_{r}^{2}|\xi_{r}|^{2}B_{0,\phi}^{2}
    \label{eq:simp}
\end{equation}
The ratio between the azimuthal and the radial component of the field to produce the same magnetic shift is $\left|\frac{\xi_{\phi}}{\xi_{r}}\right|\approx \left|\frac{k_{r}}{k_{\phi}}\right| \approx \frac{N}{\omega}$.\\
The radial component of the magnetic field has a quasi-steady component of an ``equatorial dipole" imposed at the beginning of the simulation. On top of this background field, internal gravity waves produce a highly variable component, as seen in Fig.\ref{fig:brat}. This variable component has an impact on the magnetic shift if the frequency of the probing wave is higher than the characteristic frequency at which the field oscillates. We thus consider a long time average of $|B_{\phi}|$ as representative of the azimuthal component's contribution to the magnetic shift. Fig.\ref{fig:ratios} shows that although the quasi-steady azimuthal field amplitude is higher than the radial one, the ratio between the two components is still inferior to $N/\omega$ in the whole star. We can consider the first term of \eqref{eq:simp} as dominant for the lower frequencies. \\
The considered geometry is the "equatorial dipole" such as $B_{0,r}(r,\phi) = \tilde B_{0,r}(r)\cos(\phi)$. As in a 3D case, for asymptotic g-modes, $\xi_{h}\gg\xi_{r}$ and the leading term in the expansion for $B_{r}$ not negligible compared to $B_{\phi}$ is:
\begin{equation}
    \delta \omega_{\rm mag} =   \frac{1}{4\mu_{0}\omega}\frac{\int_{0}^{R}\xi_{\phi}^*\tilde B_{0,r}(r(\xi_{\phi} \tilde B_{0,r})')'dr}{\int_{0}^{R}\xi_{\phi}^{2}r\rho dr}
\end{equation}
The prime denoting the radial derivative. Further, in a JWKB approximation, one can derive, as $\xi_{\phi}' = ik_{r}\xi_{\phi}$:
\begin{equation}
    \delta \omega_{\rm mag} =   -\frac{1}{4\mu_{0}\omega}\frac{\int_{0}^{R}\frac{|m|N^{3}\tilde B_{0,r}^{2}}{r^{3}\omega^{3}}{}\frac{1}{\rho}dr}{\int_{0}^{R}\frac{N}{|m|\omega r} dr}
\end{equation}
We retrieve:
\begin{equation}
    \delta \omega_{\rm mag} =   -\frac{m^{2}}{4\mu_{0}\omega^{3}}\frac{\int_{0}^{R}\frac{N^{3}\tilde B_{0,r}^{2}}{r^{3}}{}\frac{1}{\rho}dr}{\int_{0}^{R}\frac{N}{r} dr}
    \label{eq:shift}
\end{equation}
which is close to the expression in 3D \citep{mathisbugnet21}. We plot the asymptotic linear frequency shift obtained from Eq.\eqref{eq:shift} as a function of the mode frequency in Fig.\ref{fig:asym_shift} We see that one should expect magnetic frequency shifts below $10^{-2} \rm \mu Hz$ at the lowest frequency considered, $\sim 20 \mu Hz$. These values would be unnoticed in Fig.\ref{fig:realspacespec}.

\section{Non-dimensional Numbers}
\label{app:nond}

\begin{figure*}
    \centering
    \includegraphics[width=0.5\linewidth]{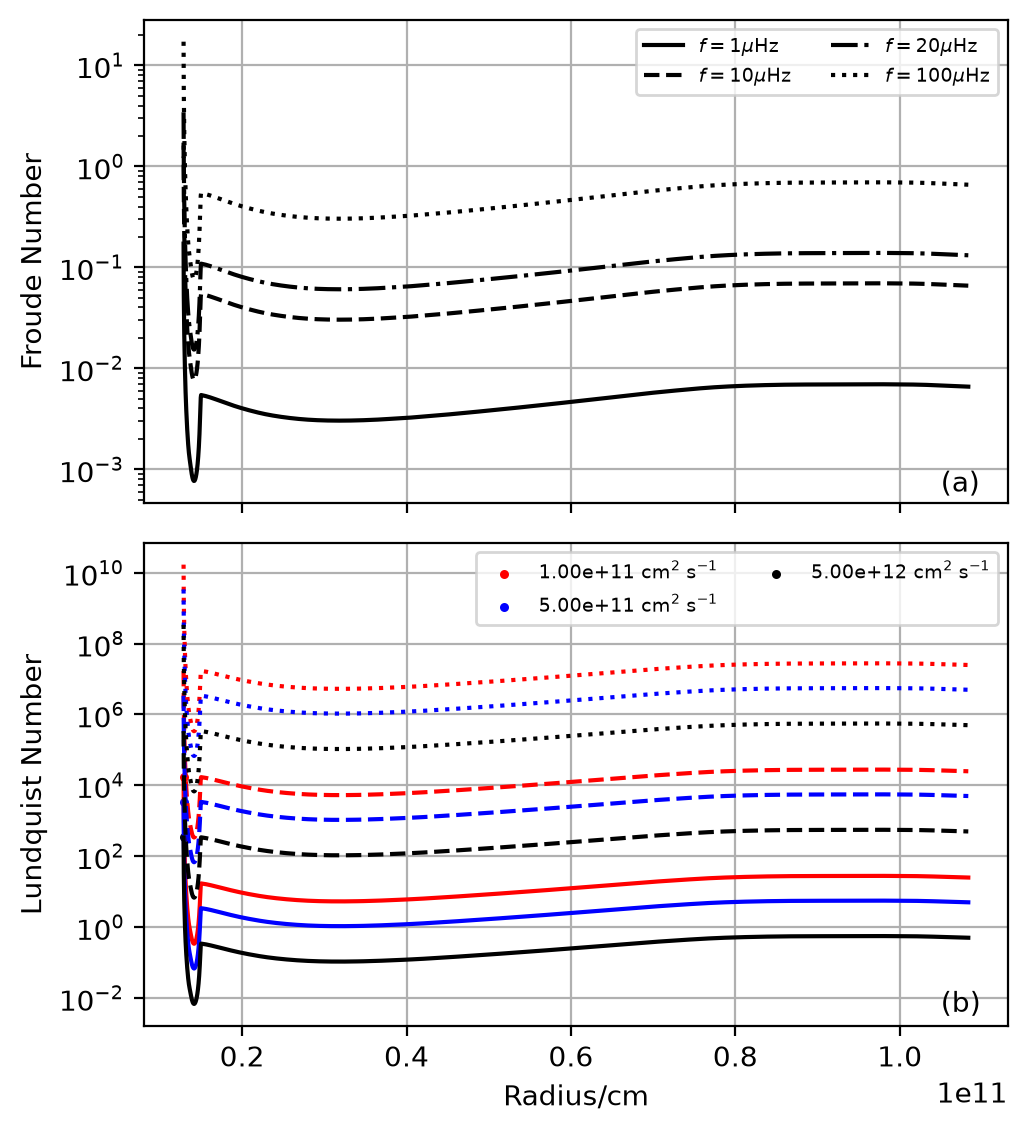}
    \caption{(a) Froude Number, $Fr = \omega/N$ for various frequency waves and (b)Lunquist Number, $Lu = \omega^3 L_{RZ}^2/(\eta N^2)$ for various frequencies and magnetic diffusivities, as functions of radius.}
    \label{fig:fr_Lu_num}
\end{figure*}

Of course, these numerical simulations have many non-dimensional parameters that are not necessarily reflective of stellar interiors.  The main parameters that are of primary importance in this study are the Reynolds number, the Froude number and the Lundquist number \citep{david2026} (magneto-buoyancy frequency is shown in the main text).  The Reynolds numbers of the simulations are shown in Table ~\ref{tab:models}, as are the Lundquist number at one position and frequency.  We show the Froude number in Fig.~\ref{fig:fr_Lu_num}(a).  Given that we use the \bvf{} frequency from the MESA model, these Froude numbers are approximately what we expect in a 2M$_{\odot}$ star at this age. If we consider a 10$\mu$Hz wave as roughly the minimum frequency we can resolve that's not dominated by thermal diffusion then, for that frequency our Froude number is around $1-4\times10^{-2}$, we note that the simulations presented in \cite{david2026} had a similar Froude number. 

\begin{figure}[!hbtp]
        \centering
        \includegraphics[trim={0.0cm 0.0cm 0 0.0cm},clip,width=0.5\linewidth]{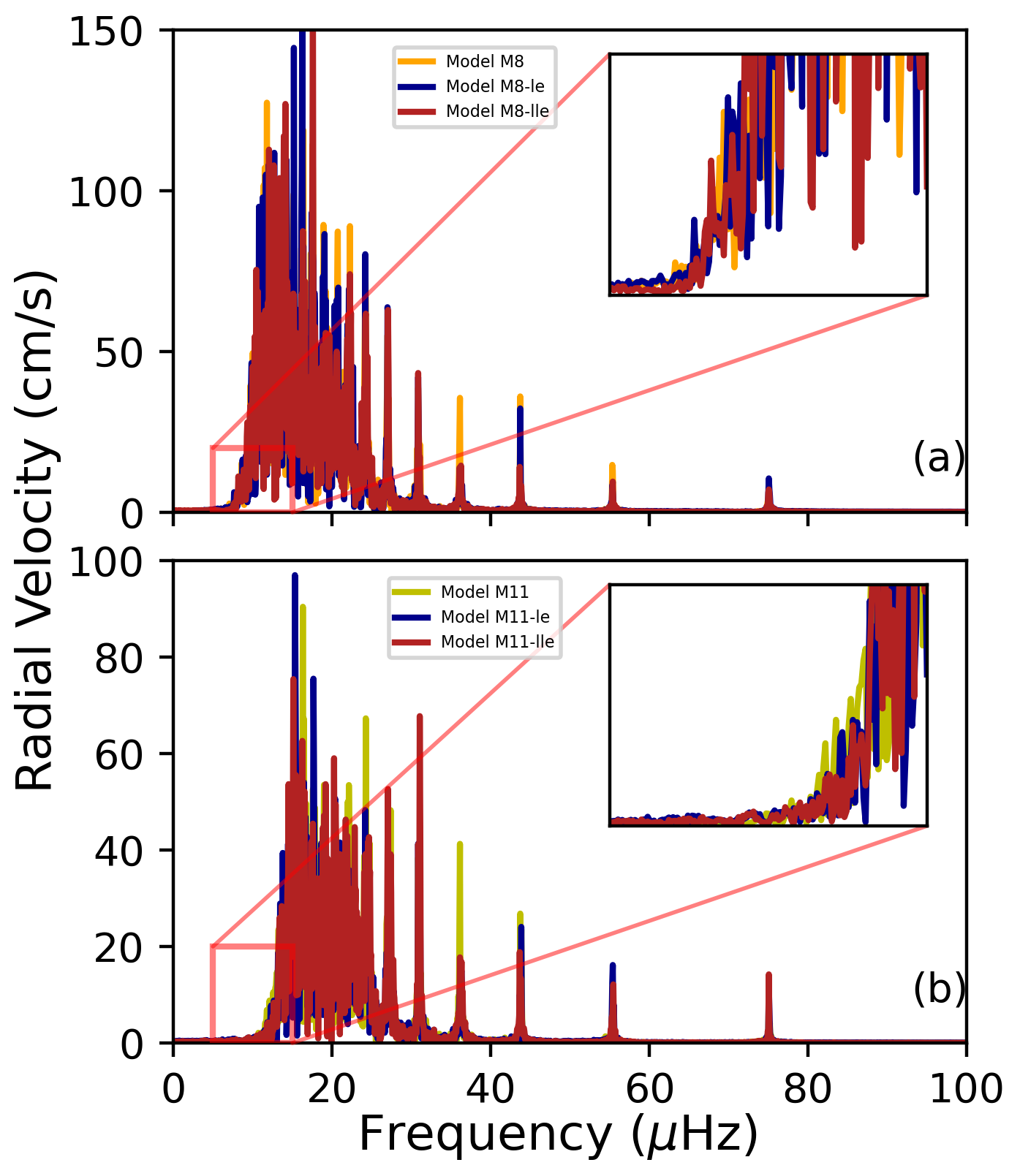}
        \caption{Comparison of models with varying magnetic diffusivity (Lundquist number).  Figure shows the Radial velocity near the top of the domain at 0.90R*. (a) Compares model M8, with magnetic diffusivities of $5\times 10^{12}$ (orange), $5\times 10{^11}$ (dark blue) and $1\times 10^{11}$ (maroon) cm$^{2}$s$^{-1}$, (b) comparing model M11, with the same diffusivities.  This results in Lundquist numbers ranging from a few hundred to several thousand at 10$\mu$Hz.}\label{fig:competa}
        \centering
\end{figure}

The Lundquist number is perhaps more relevant here.  It is a measure of the impact of magnetic dissipation on the wave mechanics, it measures the wave inertia to the wave dissipation, akin to the magnetic Reynolds number, but for a wave.  It is defined as $Lu=\omega^3L^2/\eta N^2$ \citep{david2026}.  Since our simulations have frequencies varying by orders of magnitude as well as a varying \bvf{} frequency, this number varies dramatically in our simulations, see  Fig.~\ref{fig:fr_Lu_num}(b).  The magnetic diffusivity in stellar interiors is orders of magnitude lower than that used in these simulations, resulting in much higher Lundquist numbers than we can actually simulate.  Our "global" simulations are particularly demanding requiring a higher magnetic diffusivity to capture the turbulent dynamics within the convection zone.  Since the "local" simulations do not require this, they can be run at lower magnetic diffusivities.  Therefore, we test our results by lowering the magnetic diffusivity in M8 and M11, leading to the Lundquist number changing from a few hundred for a 10$\mu$Hz wave to $\sim 10^{4}$ (depending on what radial depth one considers).  We show spectra comparing these models in Fig.\ref{fig:competa}.  There we see that the magnetic diffusivity has virtually no impact on the ability to suppress (or not) low frequency modes.  It appears that Lundquist numbers of $\sim 100$ are sufficient to capture the wave dynamics required.  We note that, for a 10$\mu$Hz wave, the Froude number is similar to that considered in \cite{david2026}.  The magneto-buoyancy ratio (square root of "$\Gamma$" in \cite{david2026}, see Fig.~\ref{fig:magnetofreq}) is, generally, substantially larger for nearly all our simulations if one is to consider the maximum field in the domain. \cite{david2026} consider a magneto-buoyancy-frequency to gravity wave frequency of approximately 0.3, meaning that their gravity wave frequency is larger than the "critical" magneto-buoyancy frequency by $\sim$ 3. As we have shown, we consider values both much higher and much lower than this critical frequency, thereby covering a range of values, as can be seen in Fig.~\ref{fig:initfields} and Fig.~\ref{fig:magnetofreq}.  


\
\end{document}